\documentclass[reprint,aps,prl,amsmath,superscriptaddress,floatfix]{revtex4-2}

\usepackage{graphicx}
\usepackage{xcolor}
\usepackage{xspace}

\begin{document}
\title{Conventional $s-$wave superconductivity and hidden peak effect\\
 in single crystals of Mo$_{8}$Ga$_{41}$ superconductor}
 
\author{Sunil~Ghimire}
\affiliation{Ames Laboratory, Ames, Iowa 50011, U.S.A.}
\affiliation{Department of Physics \& Astronomy, Iowa State University, Ames, Iowa 50011, U.S.A.}

\author{Kyuil~Cho}
\affiliation{Ames Laboratory, Ames, Iowa 50011, U.S.A.}
\altaffiliation{Department of Physics, Hope College, Holland, MI 49423, U.S.A.}

\author{Kamal~R.~Joshi}
\affiliation{Ames Laboratory, Ames, Iowa 50011, U.S.A.}
\affiliation{Department of Physics \& Astronomy, Iowa State University, Ames, Iowa 50011, U.S.A.}

\author{Makariy~A.~Tanatar}
\affiliation{Ames Laboratory, Ames, Iowa 50011, U.S.A.}
\affiliation{Department of Physics \& Astronomy, Iowa State University, Ames, Iowa 50011, U.S.A.}

\author{Zhixiang~Hu}
\affiliation{Condensed Matter Physics and Materials Science Department, Brookhaven
National Laboratory, Upton, New York 11973, U.S.A.}
\affiliation{Department of Physics and Astronomy, Stony Brook University, Stony
Brook, New York 11794-3800, U.S.A.}

\author{Cedomir~Petrovic}
\affiliation{Condensed Matter Physics and Materials Science Department, Brookhaven
National Laboratory, Upton, New York 11973, U.S.A.}
\affiliation{Department of Physics and Astronomy, Stony Brook University, Stony
Brook, New York 11794-3800, U.S.A.}
\affiliation{Shanghai Key Laboratory of Material Frontiers Research in Extreme
Environments (MFree), Shanghai Advanced Research in Physical Sciences
(SHARPS), Pudong, Shanghai 201203, China}
\affiliation{Department of Nuclear and Plasma Physics, Vinca Institute of Nuclear
Sciences, University of Belgrade, Belgrade 11001, Serbia}

\author{Ruslan~Prozorov}
\email[Corresponding author:]{prozorov@ameslab.gov}

\affiliation{Ames Laboratory, Ames, Iowa 50011, U.S.A.}
\affiliation{Department of Physics \& Astronomy, Iowa State University, Ames, Iowa 50011, U.S.A.}

\date{7 July 2024}

\begin{abstract}
London and Campbell penetration depths were measured in single crystals of the endohedral gallide cluster superconductor, Mo$_{8}$Ga$_{41}$. The full temperature range superfluid density is consistent with the clean isotropic $s-$wave weak-coupling BCS theory without any signs of the second gap or strong coupling. The temperature dependence of the Campbell length is hysteretic between zero-field cooling (ZFC) and field-cooling (FC) protocols, indicating an anharmonic vortex pinning potential. The field dependence of the effective critical current density, $j_{c}\left(H\right)$, reveals an unusual result. While in the ZFC protocol, $j_{c}\left(H\right)$ is monotonically suppressed by the magnetic field, it exhibits a profound ``hidden'' peak effect in the FC protocol, that is, without a vortex density gradient. We suggest a possible novel mechanism for the formation of the peak effect, which involves both static and dynamic aspects. 
\end{abstract}
\maketitle

\section{Introduction}

Unconventional superconductivity is a perpetual theme in current research. Particular attention is devoted to multi-gap superconductivity \cite{Crabtree2003,Kresin2013}. Although it was theoretically introduced in 1959 \cite{Suhl1959,Moskalenko1959} and later received experimental confirmation \cite{Binnig1980}, it has not become a part of the mainstream research effort. All changed with the discovery of two-gap superconductivity in MgB$_{2}$ in 2001 \cite{Nagamatsu2001}, after which multi-gap physics has become one of the most studied phenomena in modern superconductivity \cite{Crabtree2003,Kresin2013}. Inspired by MgB$_{2}$, many binary systems have been investigated, and endohedral gallide cluster compounds are among the actively researched families of materials \cite{sciencedirect_1984,Xie2015,Sirohi2019,Verchenko2021}. 

In this paper, we study one of its members, Mo$_{8}$Ga$_{41}$, with superconducting transition temperature $T_{c}=9.7\:\text{K}$ and an upper critical field $H_{c2}\approx8.3-8.7\:\text{T}$ \cite{sciencedirect_1984,Verchenko2020,Verchenko2021}. Although this compound has been known for more than forty years \cite{sciencedirect_1984}, it has attracted recent attention for possible deviations from the conventional single-gap $s-$wave weak-coupling Bardeen-Cooper-Schrieffer(BCS) superconductivity \cite{Bardeen1957a,Bardeen1957}. A quantum oscillations study of Mo$_{8}$Ga$_{41}$ inferred three-dimensional electronic bands with strong coupling to phonons \cite{Petrovic_QO_2020}. Specific heat measurements reported larger than weak coupling jump at $T_{c}$ \cite{Verchenko2016}. A scanning tunneling microscopy (STM) study of Mo$_{8}$Ga$_{41}$ crystals identified two gaps in the density of states \cite{Sirohi2019}. However, their larger gap gives $\Delta/T_{c}=1.857$, which is hardly possible to separate from the weak-coupling BCS value, $\Delta/T_{c}=1.764$ and its temperature dependence follows the BCS in the full temperature range. A study of muon spin relaxation ($\mu\text{SR}$) concluded that the temperature-dependent magnetic penetration depth can be described by two $s-$wave superconducting gaps \cite{Verchenko2017}. However, the larger gap to $T_{c}$ ratio gives an unphysical result, $\Delta/T_{c}=1.163 < 1.764$, probably because these measurements were carried out in an external dc magnetic field when Abrikosov vortices are present, making it difficult to identify the pure London penetration depth. Notably, their data of $\lambda^{-2}\left(T\right)$ can be reproduced by a single-gap weak-coupling BCS model with much more reasonable, $\Delta/T_{c}=2.1$. Another study of the temperature dependence of the lower critical magnetic field, $H_{c1}$, measured using a miniature Hall probe array, showed that the results can be fitted well by both single-gap and two-gap superconductivity with minor differences \cite{Marcin_2021}. Furthermore, a combined study using ac calorimetry and STM on the same high-quality crystals showed only one intrinsic gap and some traces of additional superconducting phases, which could be detected as the second gap in the STM measurements \cite{Marcin2019}. To conclude, there is a significant disagreement about the nature of superconductivity in Mo$_{8}$Ga$_{41}$ with secondary phases and/or surface layers potentially ``contaminating'' the results. Therefore, further investigation is needed.

With regard to vortex properties, there is only very limited literature. Magnetic measurements of polycrystalline samples showed conventional-looking hysteresis $M\left(H\right)$ loops with the specific power-law magnetic field dependence of irreversible magnetization, indicative of a strong pinning \cite{vanderBeek2002,Beek2012a,Willa2015,Willa2015a,Willa2016,Willa2021}. The persistent current density in zero field was estimated from the Bean model \cite{Bean1964,Bean1962}, $j_{p}\left(T=2\:\text{K}\right)=0.3\:\text{MA/cm}^{2}$ \cite{Neha2019}. Another study of single crystals using miniature Hall probe arrays estimated $j_{p}\left(T=2\:\text{K}\right)=0.016\:\text{MA/cm}^{2}$ from the Maxwell equation \cite{Marcin2020}. In principle, such a difference can be attributed to the dissimilarity between the polycrystalline and single-crystalline samples. The latter study confirmed the strong-pinning scenario and suggested single-gap superconductivity, perhaps contaminated by secondary phases. 

One of the most interesting mixed-state features is the non-monotonic dependence of the irreversible component of magnetization on a magnetic field or temperature. Depending on the context and author's preferences, this feature can be called the ``peak effect'', the ``second magnetization peak'' or the ``fishtail'' \cite{DeSorbo1964,Larkin1979,Daeumling1990,KrusinElbaum1992,Klein1994,Blatter1994,Brandt_1995,Tang1996,Giller1997,Banerjee2000,Mikitik2001,Prozorov2008}. Since any measurement has a certain experimental time window, the measured magnetic moment and hence the persistent current density $j_p$, are affected by magnetic relaxation, which is exponentially fast at current densities close to the critical current $j_c>j_p$ \cite{Blatter1994,Brandt_1995,Yeshurun1996}. Consequently, there is an ongoing debate about the static or dynamic origin of the peak effect. The ``static'' explanation suggests actual non-monotonic behavior of the unrelaxed critical current, $j_{c}\left(H\right)$, which would imply an unusual pinning mechanism, for example, due to the softening of the vortex lattice at low fields and close to $H_{c2}$ \cite{Pippard1969,Larkin1979}, two different vortex phases \cite{Konczykowski2000}, or a crossover from collective to plastic creep mechanism \cite{Abulafia1996,Giller1997}. The ``dynamic'' explanation involves a field-dependent magnetic relaxation that is faster at low magnetic fields, for example, in the weak collective pinning and creep model \cite{Blatter1994}. 

In this work, we address both the superconducting gap structure probed by measuring the London penetration depth and the theoretical critical current density probed by measuring the Campbell penetration depth \cite{Campbell_1969,Campbell_1971,Koshelev1991,Willa2016} in Mo$_{8}$Ga$_{41}$ single crystals. Although the superfluid density is well described by the isotropic single-gap weak-coupling BCS theory, the vortex behavior is unusual. We found an unexpected ``hidden'' peak effect in the ``true'' $j_{c}\left(H\right)$ in the field-cooling protocol, when persistent current density is zero (no vortex density gradient) which is inaccessible to other types of magnetization measurements, global or local. We note that from many superconductors in which we measured Campbell length, there is only one example, LiFeAs, where we observed similar behavior, namely, monotonic in field ZFC current density and non-monotonic peak effect in a FC protocol \cite{Propmann_PRB_2011}.

\section{Samples and Methods}

\textbf{Samples:} Single crystals of Mo$_{8}$Ga$_{41}$ were grown by the high-temperature self-flux method. Mo and Ga were mixed in an 8: 500 ratio in an alumina crucible and sealed in an evacuated quartz tube. The ampule was heated up to 850$^{0}$C in two hours, held at 850$^{0}$C for 10 hours, and then slowly cooled to 170$^{0}$C for 55 hours, when the crystals were decanted \cite{Petrovic_QO_2020}.

\textbf{Lower critical field:} The lower critical field, $H_{c1}$, was measured using local optical magnetometry based on nitrogen vacancy centers in diamond (NV$^{-}$centers) \cite{Nusran2018,Joshi2019,Joshi2020}. Due to the specifics of the NV-center low-energy spectrum, the stimulated fluorescence amplitude depends on the applied magnetic due to Zeeman levels splitting. If the magnetic field is applied along the $\hat{z}$ direction, this
results in two peaks in the optically detected magnetic resonance (ODMR) with their splitting, $\Delta f=2g_{NV}B/\sqrt{3}$, where $g_{NV}=2.8\:\text{MHz/Oe}$ is the NV center gyromagnetic ratio and $\sqrt{3}$ in the denominator takes into account the possible projections of the applied magnetic field on the $N-V$ bond direction in our {[}100{]} oriented diamond crystalline film. The NV centers are implanted 20 nm below the surface and the diamond film is placed on top of the flat sample. Measurements are performed as close to the edge of the sample as possible. This is similar to the micro-Hall probe measurements mentioned in the Introduction \cite{Marcin_2021}. 

\textbf{London penetration depths:} The London penetration depth, $\lambda_{L}(T)$, was measured using a sensitive frequency-domain self-oscillating tunnel-diode resonator (TDR) operating at a frequency around 14~MHz. The measurements were performed in a $^{3}$He cryostat with the base temperature of about $T_{min}=0.4\:\text{K}$, which is $0.041T_{c}$, which gives us enough range to examine the low-temperature limit, which starts below $T_{c}/3$, where the superconducting gap is approximately constant. The experimental setup, measurement protocols, and calibration are described in detail elsewhere \cite{VanDegrift1975,Prozorov2000,Kim2018,Prozorov2021}. 

In the experiment, the temperature dependence of the relative resonant frequency shift $\delta f=f\left(T\right)-f\left(T_{min}\right)$ is measured. This quantity is proportional to the magnetic susceptibility of the sample, which, in turn, is proportional to $\delta f/f_{0}=G\Delta\lambda\left(T\right)/R$ where $R$ is the effective dimension of the sample calculated for each particular geometry and $G$ is the dimensionless calibration constant \cite{Prozorov2021}. The parameter $f_{0}=14\:\text{MHz}$ is the frequency of the empty resonator. The calibration constant $G$ is measured by physically pulling the sample out of the coil at $T_{min}$. The total London penetration depth is obtained as $\lambda\left(T\right)=\lambda\left(0\right)+\Delta\lambda(T)$, where $\Delta\lambda(T)=\lambda(T)-\lambda(0.4\:\text{K})$ is the measured change in the penetration depth as described above and the absolute value $\lambda(0)$ is estimated separately using the NV-centers optical magnetometry. Finally, the normalized superfluid density that can be directly compared with the theory is evaluated as $\rho_{s}=\left(\lambda\left(0\right)/\lambda\left(T\right)\right)^{2}$.

\textbf{Campbell penetration depths:} The Campbell penetration depth, $\lambda_{C}(T)$, is measured exactly the same way as the London penetration depth, but in a finite applied magnetic field, which produces Abrikosov vortices in the sample. Then, the measured penetration depth, $\lambda_{m}$, has two contributions, the usual London penetration depth, $\lambda$, and the Campbell penetration depth $\lambda_{C}$, which is a characteristic length scale over which a small ac perturbation is transmitted elastically by the vortex lattice into the sample \cite{Campbell_1969,Campbell_1971,Gaggioli2022,Willa2015a,Willa2015}. More specifically, the amplitude of the ac perturbation must be small enough so that the vortices remain in their potential well, and their motion is described by the reversible linear elastic response. In this case, $\lambda_{m}^{2}=\lambda^{2}+\lambda_{C}^{2}$ \cite{Brandt1991,Koshelev1991}. This requirement of a very small amplitude makes most conventional ac susceptibility techniques inapplicable to Campbell length measurements. In our case, the excitation magnetic field is approximately 20 mOe, which is surely well below $H_{c1}$ for most of the temperature range. It is important to note that conventional ac and dc measurements, where displacement of vortices out of their potential wells is involved, probe the Bean persistent current density \cite{Blatter1994,Brandt_1995}, whereas Campbell length measurements probe the curvature of the effective pinning potential \cite{Brandt1991,Willa2015,Willa2015a}. This information is inaccessible for conventional measurements.

\begin{figure}[tb]
\includegraphics[width=8cm]{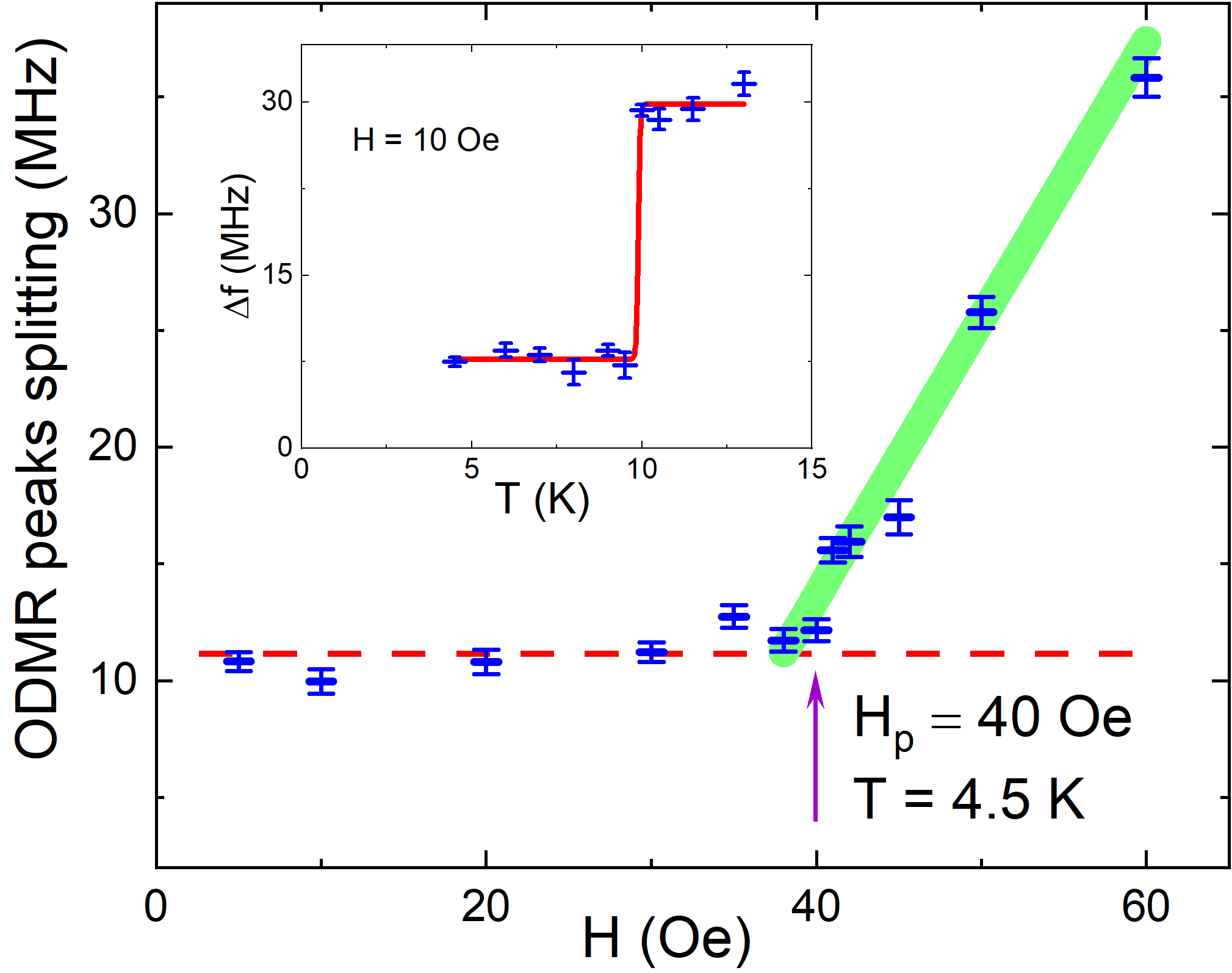} 
\caption{Determining the magnetic field of the first flux penetration, $H_{p}$, in Mo$_{8}$Ga$_{41}$ at $T=4.5\:\text{K}$ by local optical magnetometry using NV-centers. Zeeman splitting is measured at the edge of the superconductor after zero field cooling. A sharp departure from a constant value occurs at $H_{p}=40\:\text{Oe}$ when vortices start penetrating the sample. The inset shows sharp superconducting transition measured at the same spot.}
\label{fig:NV}
\end{figure}

\section{Results and Discussion}

\subsection{The absolute value of $\lambda_{L}$(0) from NV-centers optical magnetometry}

Figure~\ref{fig:NV} shows the ODMR splitting near the edge of the sample as a function of the applied dc magnetic field at $T=4.5\:\text{K}$. A sharp break from a constant value occurs in the magnetic field of the first flux penetration $H_{p}\left(4.5\:\text{K}\right)=40\:\text{Oe}$. The inset shows a superconducting transition measured at the same spot. Using a revised effective demagnetization factor for a $2a \times 2b \times 2c$ cuboid, $N^{-1}=1+\frac{3c}{4a}(1+\frac{a}{b})$ \cite{Prozorov2018}, the true $H_{c1}\left(4.5\:\text{K}\right)=85\:\text{Oe}$ is obtained. The absolute value of the London penetration depth is then estimated by solving $H_{c1}=\frac{\phi_{0}}{4\pi\lambda^{2}}\left(\ln\frac{\lambda}{\xi}+0.497\right)$ \cite{Tinkham2004}, where $\xi=\sqrt{\phi_{0}/2\pi H_{c2}}$ is the coherence length and $\phi_{0}$ is the magnetic flux quantum. The upper critical field at the temperature of interest, $H_{c2}\left(4.5\:\text{K}\right)=5.25\:\text{T}$, was obtained from the specific heat and magnetization measurements of high-quality crystals \cite{Verchenko2016}, which gives $\xi\left(4.5\:\text{K}\right)=8\:\text{nm}$. From this we obtain $\lambda\left(4.5\:\text{K}\right)=281\:\text{nm}$. Finally, using a known analytic approximation of $\lambda\left(T\right)$ for an $s-$wave weak-coupling superconductor \cite{Prozorov2006}, we can evaluate $\lambda\left(0\right)=\lambda\left(t\right)\sqrt{1-t^{4}}$, resulting in $\lambda\left(0\right)=274\:\text{nm}$. This value will be used to construct the superfluid density.

\subsection{London penetration depth and superfluid density}

\begin{figure}[tb]
\includegraphics[width=8cm]{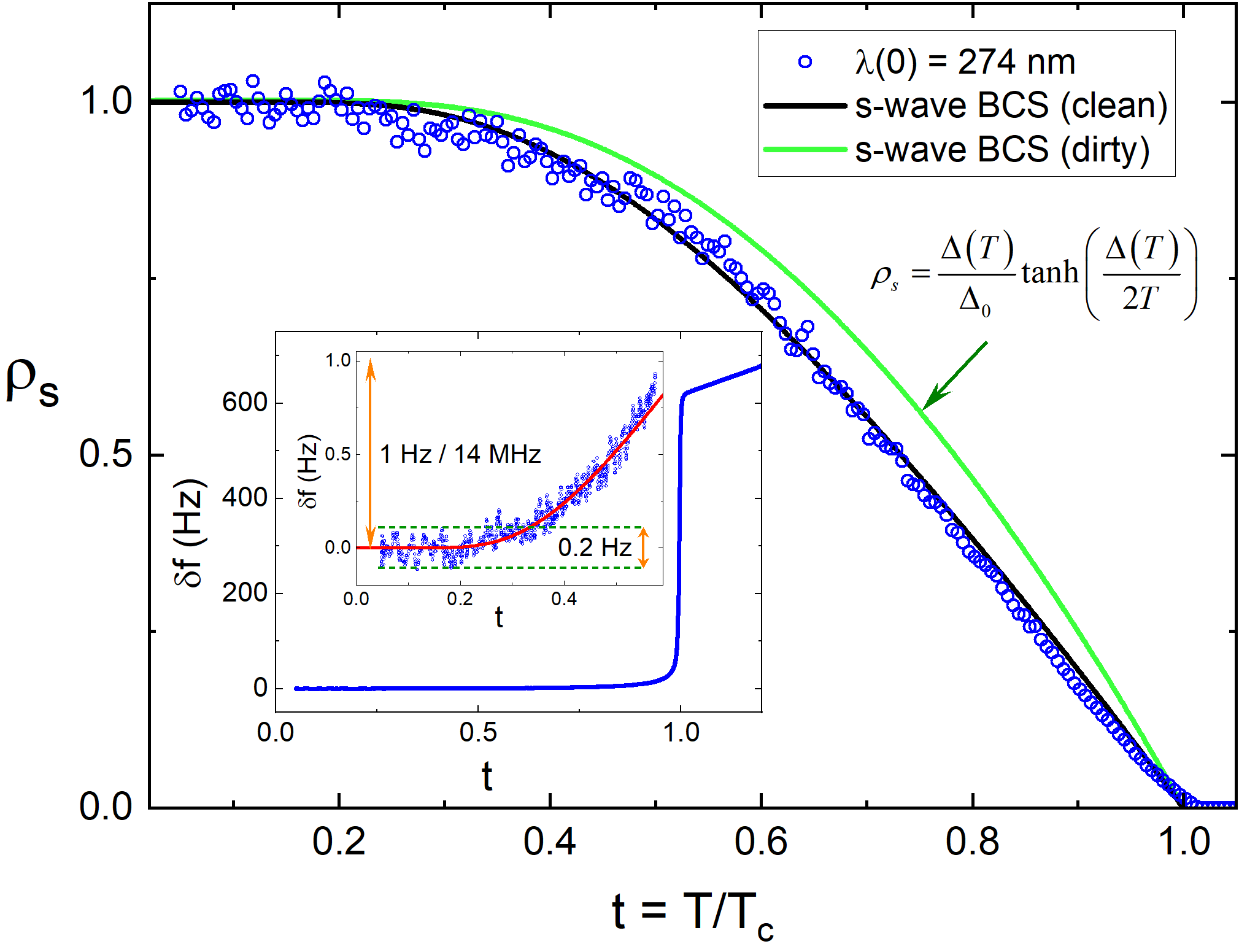} 
\caption{Main panel: superfluid density, $\rho_{s}(T)\equiv\lambda^{2}(0)/\lambda^{2}(T)=(1+\Delta\lambda(T)/\lambda(0))^{-2}$ calculated using the absolute value of London penetration depth, $\lambda\left(0\right)=274\:\text{nm}$, and the temperature-dependent variation, $\Delta \lambda(T)$ measured using a tunnel diode resonator. The black line shows a theoretical curve (not a fit!) for the isotropic $s-$wave weak-coupling BCS superconductor. The green curve shows the numerical dirty limit, well approximated analytically by $\rho_{s}=\left(\Delta(T)/\Delta(0)\right)\tanh\left(\Delta(T)/2T\right)$. The outer inset shows the raw data, $\delta f\left(t\right)$, in the full temperature range. The inner inset shows low-temperature zoom with the data shown by blue circles and low-temperature BCS asymptotic shown by the red line.}
\label{fig:SFD}
\end{figure}

We are now ready to calculate the normalized superfluid density, $\rho_{s}(T)\equiv\lambda^{2}(0)/\lambda^{2}(T)=(1+\Delta\lambda(T)/\lambda(0))^{-2}$. The result is shown in Fig.\ref{fig:SFD}. The raw data, the resonator frequency shift, $\delta f\left(T\right)$, in a full temperature range are shown in the inset and the low temperature range is shown in the inner inset. We present the raw data to demonstrate that some apparent noise is not due to measurement problems but to a very small signal. The inner inset of Fig.\ref{fig:SFD} shows that the total frequency change when the temperature increases from $T_{min}$ to $0.5T_{c}$ is only $1\:\text{Hz}$. The noise level indicated by green dashed lines is $0.2\:\text{Hz}$, which means that we have a 10 parts per billion accuracy, which is very good. The red line in the inner inset is the isotropic asymptotic, $\delta f(T)=A\sqrt{\pi\delta/2t}e^{-\delta/t}$ with a fixed ratio $\delta=\Delta(0)/T_{c}\approx1.76$ leaving only one free scaling parameter, $A$. This equation is applicable to $\delta f$, because, as described in Section II, $\delta f\sim\Delta\lambda$. 

Next we calculated the normalized superfluid density using $\lambda(0)=274\:\text{nm}$ determined using NV optical magnetometry as described in Section III.B. The main panel of Fig.\ref{fig:SFD} shows the data (blue circles) and two fixed curves (not a fit!) for clean (black) and dirty (green) limits of the isotropic $s-$wave weak-coupling BCS superconductor. Theoretical curves were computed numerically using a self-consistent solution of the Eilenberger equations \cite{Eilenberger1968,Prozorov2011}. We note that the non-magnetic scattering dirty limit can be described analytically, $\rho_{s}=\left(\Delta(T)/\Delta(0)\right)\tanh\left(\Delta(T)/2T\right)$ \cite{Kogan2013}. Clearly, the clean limit fits the very data well in the entire temperature range. There is no indication of a multi-gap behavior, which usually appears as a convex curvature of $\rho_{s}\left(T\right)$ at elevated temperatures where the smaller gap is coupled to a larger gap by proximity \cite{Prozorov2006,Prozorov2011}. 

\begin{figure}[tb]
\includegraphics[width=8cm]{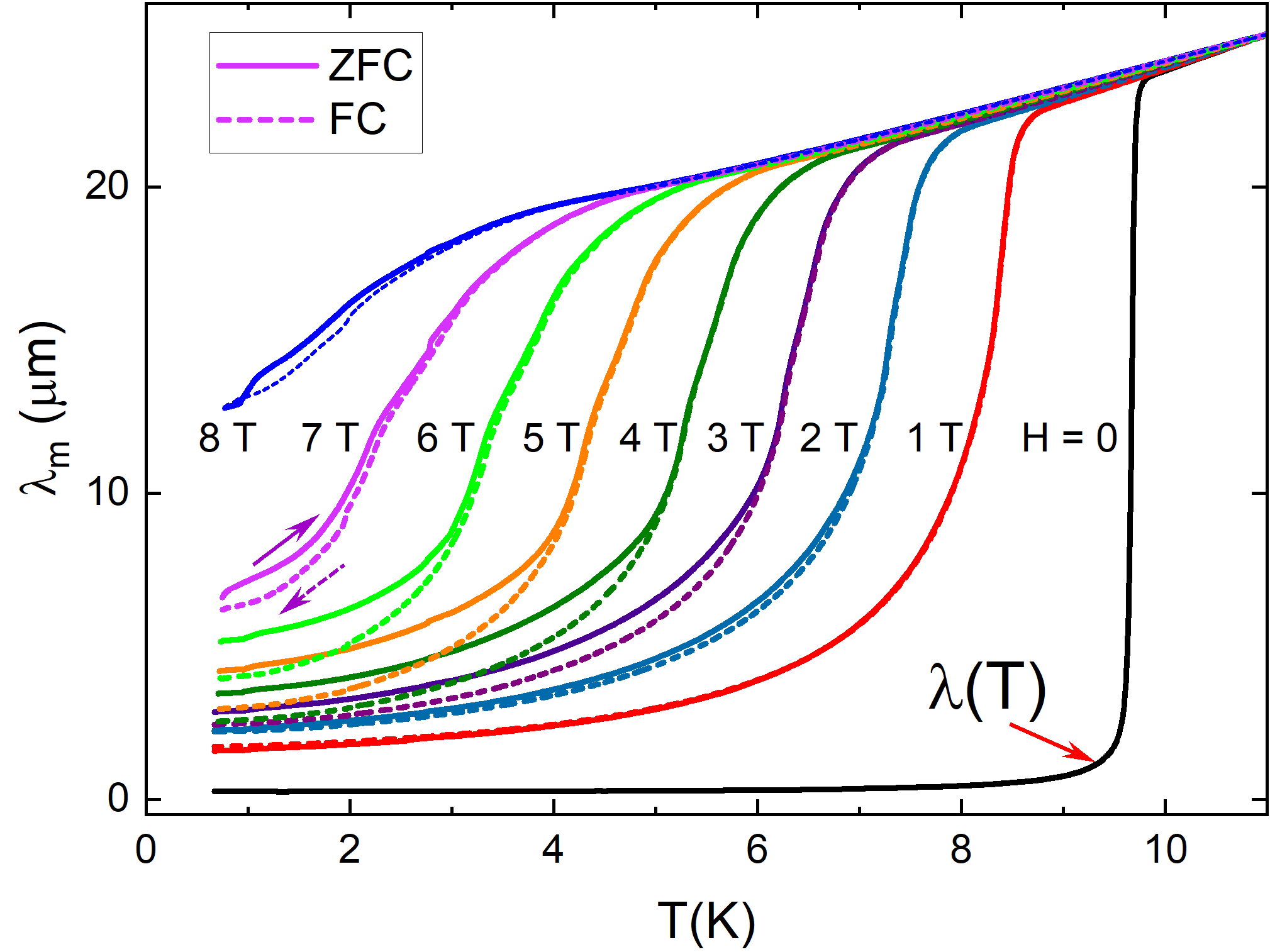} 
\caption{Total magnetic penetration depth, $\lambda_{m}(T)\lambda(0)+\Delta\lambda(T)$, as a function of temperature, measured on warming (ZFC, solid lines) and cooling (FC, dashed lines) at different applied dc magnetic fields indicated next to each curve. Two arrows near a 7 T curve indicate the direction of the temperature sweep.}
\label{fig:Lm}
\end{figure}

\subsection{Campbell penetration depth and critical current density}

Figure~\ref{fig:Lm} shows the total magnetic penetration depth, $\lambda_{m}$, measured upon warming (solid curves) after zero-field cooling (ZFC, dashed curves)and on cooling from above $T_{c}$ (FC) at different applied magnetic fields, shown in the figure. The hysteretic behavior between ZFC and FC protocols is not due to vortex density gradient as occurs in dc magnetization measurements. This hysteresis comes from a non-parabolic vortex pinning potential, $U(r)$. Note that the measured $\Delta\lambda_{m}(T)$ saturate, approaching the normal state. Of course, the actual penetration depth diverges at $T\rightarrow T_{c}\left(H\right)$, but in the normal state, $\lambda_{m}$ cannot exceed the skin depth, $\delta_{\text{skin}}=\sqrt{\rho/\mu_{0}\pi f}$, where $\mu_{0}=4\pi\times10^{-7},\text{H/m}$ is the vacuum permeability and $\rho$ is the resistivity just above $T_c$. Therefore, the saturation value in Figure~\ref{fig:Lm} is temperature dependent through $\rho\left(T\right)$. In fact, such measurements can be used for contactless resistivity measurements \cite{Prozorov2007}.

\begin{figure}[tb]
\includegraphics[width=8cm]{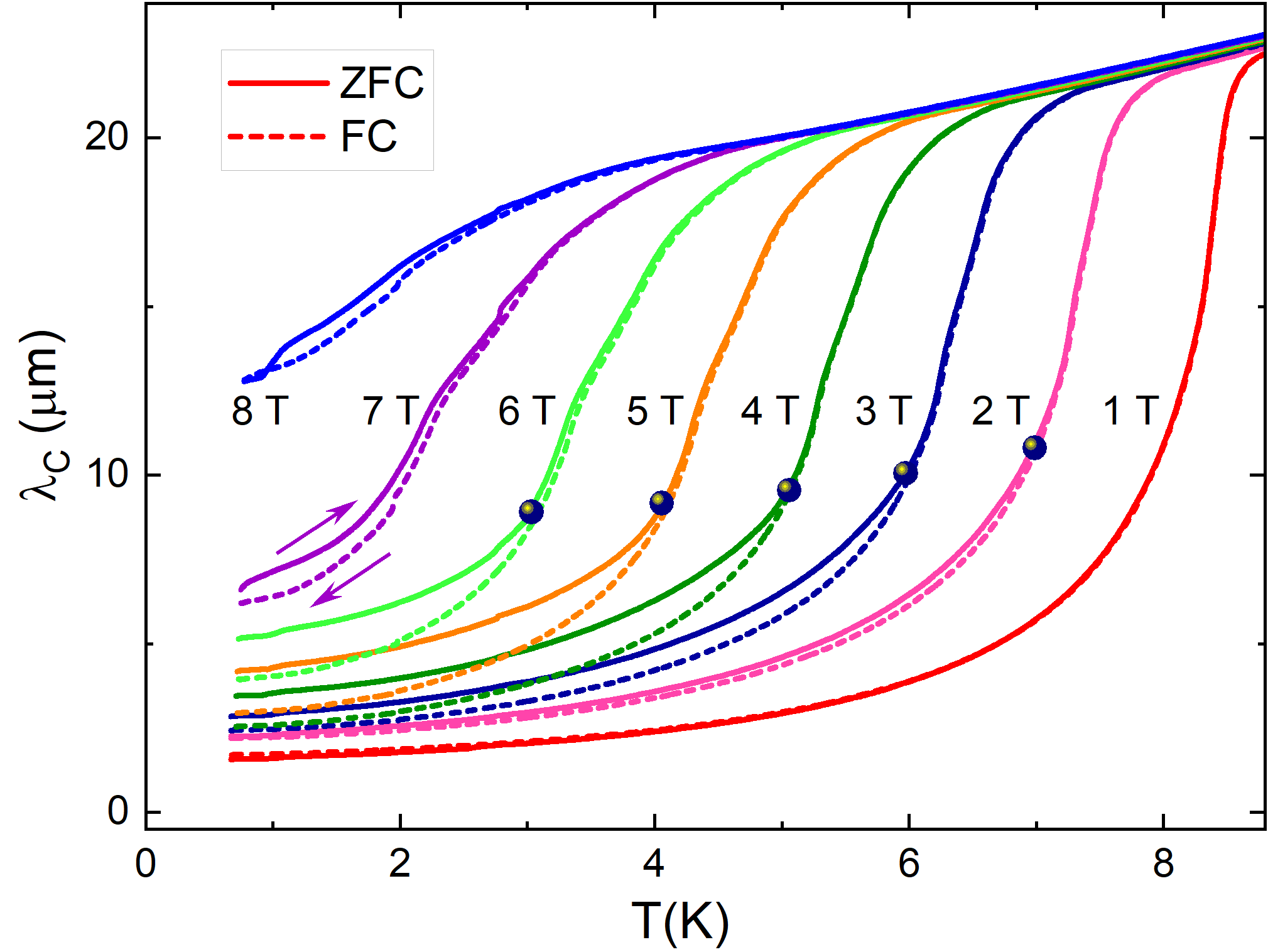} 
\caption{Campbell penetration depth, $\lambda_{C}=\sqrt{\lambda_{m}^{2}-\lambda_{L}^{2}}$, as a function of temperature measured on warming (ZFC, solid lines) and cooling (FC, dashed lines) in different dc magnetic fields, shown next to each curve. The arrows indicate the direction on one of the curves.}
\label{fig:Lc(T)}
\end{figure}

\begin{figure}[tb]
\includegraphics[width=8cm]{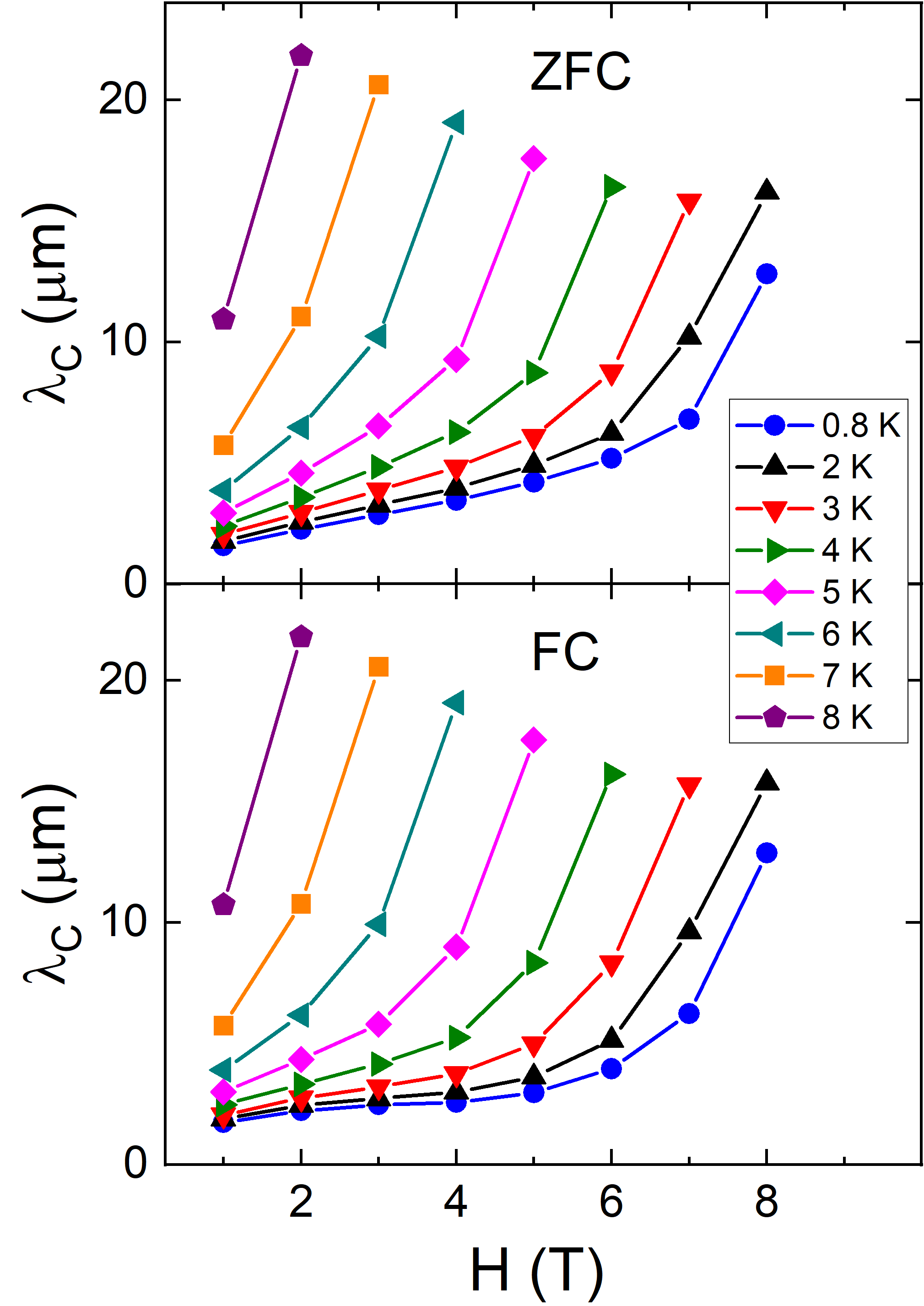} 
\caption{Campbell length, $\lambda_{C}$, as a function of a magnetic field at fixed temperatures as shown in the legend. The symbols mark the points obtained from the isothermal slices of Fig.\ref{fig:Lc(T)}.}
\label{fig:Lc(B)}
\end{figure}

We now evaluate the Campbell length, $\lambda_{C}=\sqrt{\lambda_{m}^{2}-\lambda_{L}^{2}}$. Figure \ref{fig:Lc(T)} shows the calculated $\lambda_{C}\left(T\right)$ with the same type and color of the curves for the indicated magnetic fields as in Fig.\ref{fig:Lm}. With an increasing magnetic field, the $\lambda_{C}\left(T\right)$ curves move upward, indicating the field dependence, which can be determined from Fig.\ref{fig:Lc(T)} by taking isothermal slices. The result is shown in \ref{fig:Lc(B)} where the top panel shows the ZFC $j_c(B)$ curves, while the bottom panel shows the FC curves. 

In the original Campbell model \cite{Campbell_1969,Campbell_1971}, $\lambda_{C}^{2}=\phi_{0}H/\alpha$ and $j_{c}=\alpha r_{p}/\phi_{0}=Hr_{p}/\lambda_{C}^{2}$, where $r_{p}$ is the radius of the pinning potential, usually assumed to be approximately equal to the coherence length, $\xi$, and $\alpha$ is the so-called Labusch constant, the curvature of the pinning potential, $\alpha=d^{2}U/dr^{2}$. Note that we used SI units with $H$ being the magnetic field strength in A/m. When magnetic induction $B$ is used in tesla, the formulas must replace $H=B/\mu_{0}$.

Generally, the Campbell length $\lambda_{C}(H)\propto H^{\beta}$ with $\beta=0.5$ in the original Campbell theory, which was observed in the systems previously studied \cite{Prozorov2004,Kim2013}. However, other superconductors show a different exponent $\beta$. For example, a more concave $\lambda_{C}(H)$ with $\beta=0.25$ was observed in the low carrier density superconductor YPtBi~\cite{hynshoo_2021}. In the present case of Mo$_{8}$Ga$_{41}$, a very different behavior with a field-dependent exponent, $\beta\left(B\right)>0.5$, indicating that the pinning potential is anharmonic with a field-dependent Labusch parameter. There is an important difference between the ZFC and FC protocols. While the former probes the Labusch parameter closer to the edge of the pinning potential where vortices are biased by the Bean persistent current, $j_p$, the FC measurements probe the theoretical critical current with vortices oscillating at the bottom of the pinning potential \cite{Prozorov2003}. This is a good proxy for the true critical current density since the difference between the ZFC and FC branches is not large and the degree of anharmonicity of the pinning potential is small.  

At low fields and temperatures, the FC curves, $\lambda_{C}(H)$, start with the exponent $\beta$ close to $0.5$, but then change to a much larger value. This is likely due to a crossover from a single vortex pinning to the collective pinning of vortex bundles \cite{Blatter1994}.

\begin{figure}[tb]
\includegraphics[width=8cm]{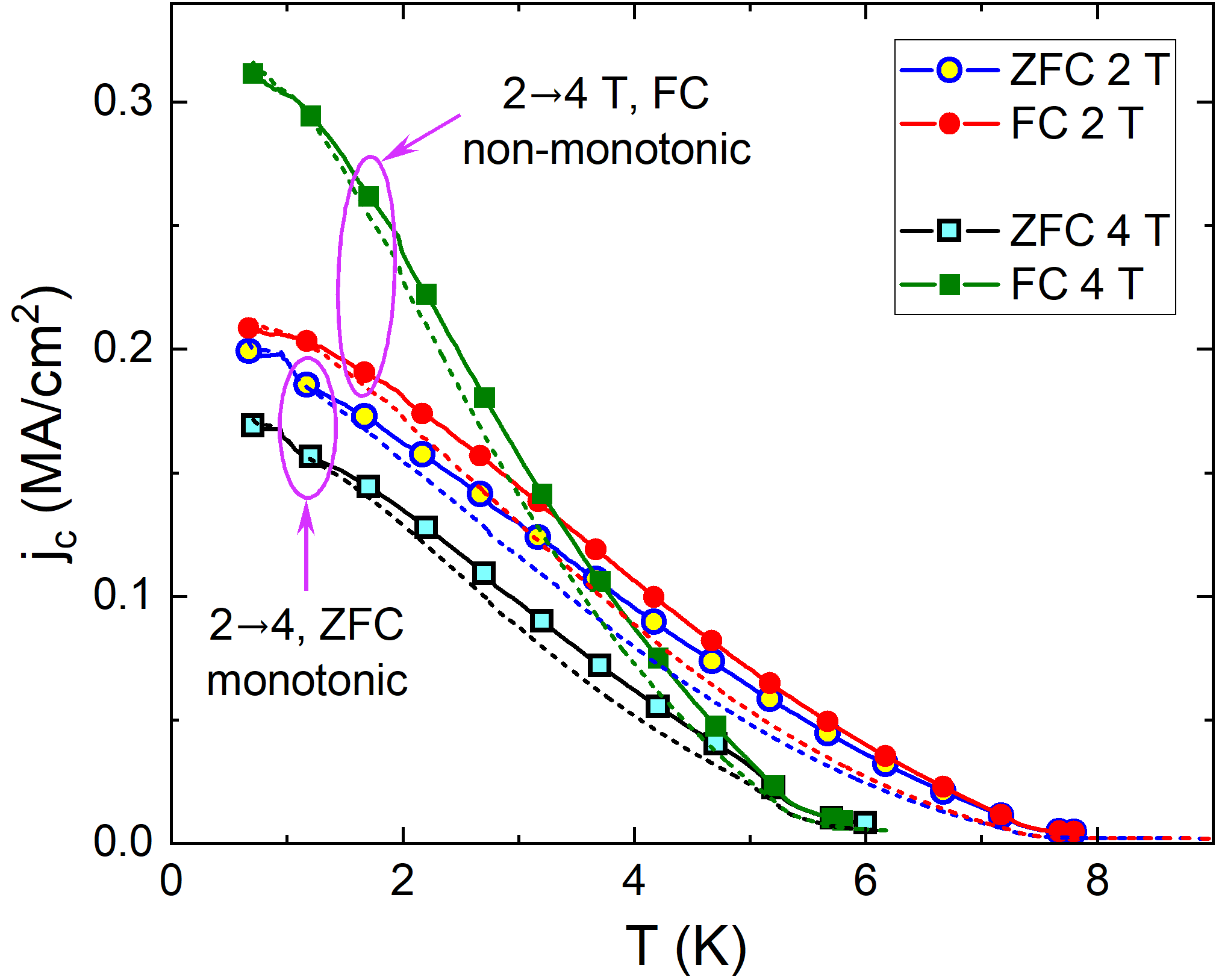} 
\caption{Temperature dependence of the theoretical critical current density in $2\:\text{T}$ (circles, empty - ZFC, filled - FC) and $4\:\text{T}$ (squares) evaluated from $j_{c}=Hr_{p}/\lambda_{C}^{2}$ using the data shown in Fig.\ref{fig:Lc(T)}. The dashed lines are calculated with a constant $r_{p}=\xi\left(0\right)=6.4\:\text{nm}$. Ellipses group the data into ZFC and FC protocols. The critical current decreases with an increasing magnetic field in the entire temperature interval in a ZFC protocol $j_{c}\left(2\:\text{T}\right)>j_{c}\left(4\:\text{T}\right)$, but the trend is opposite in the case of a FC protocol.}
\label{fig:Jc(T)}
\end{figure}

Figure~\ref{fig:Jc(T)} shows the theoretical critical current density evaluated from the Campbell model using the data from Fig.\ref{fig:Lc(T)}, $j_{c}=r_{p}H/\lambda_{C}^{2}$ \cite{Campbell_1969,Campbell_1971}. The range of the pinning potential, $r_{p}$, is usually associated with the coherence length, which does not change much over a large temperature interval. To be thorough, we extracted the temperature-dependent $\xi\left(T\right)$ from the experimental upper critical field \cite{Verchenko2016}, and found that it only doubles at $T=7.5\:\text{K}$. To facilitate the comparison and avoid overcrowding, Fig.~\ref{fig:Jc(T)} shows the theoretical critical current density as a function of temperature for two values of the applied magnetic field, $2\:\text{T}$ and $4\:\text{T}$. Symbols show the results with a temperature-dependent $\xi\left(T\right)$, while dashed lines assume a fixed $r_{p}=\xi\left(0\right)=6.4\:\text{nm}$. Expectedly, the difference increases for larger temperatures, but it is still minor and does not change the overall functional form. In ZFC measurements, the curves at $4\:\text{T}$ lie below those of $4\:\text{T}$, so that $j_{c}\left(2\:\text{T}\right)>j_{c}\left(4\:\text{T}\right)$, showing an expected monotonic decrease of $j_{c}$ with a magnetic field. Surprisingly, the ZFC curves reverse this order, so $j_{c}\left(2\:\text{T}\right)<j_{c}\left(4\:\text{T}\right)$, showing a clear increase of $j_{c}$ with the increasing magnetic field.

\begin{figure}[tb]
\includegraphics[width=8cm]{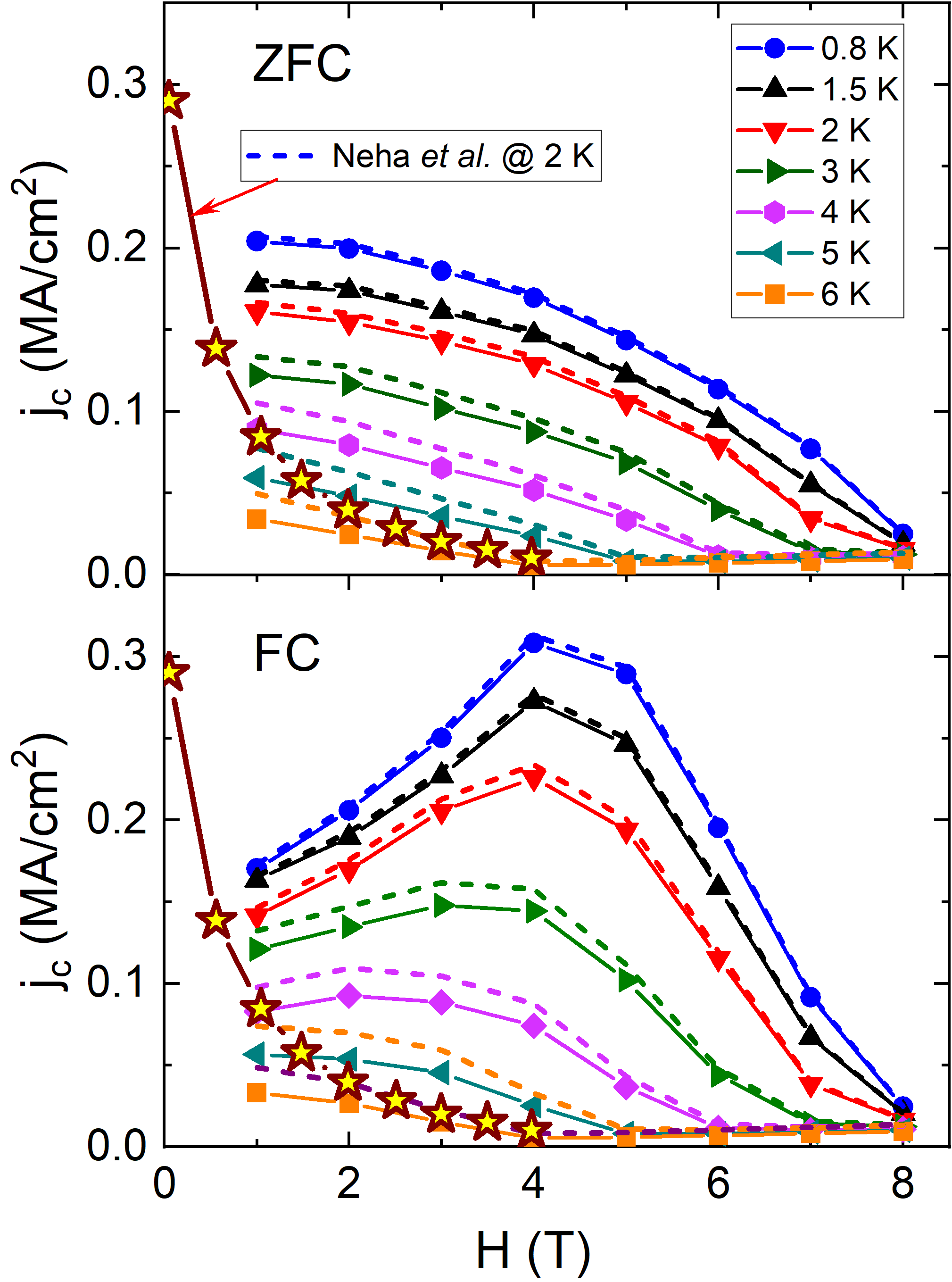} 
\caption{ Campbell penetration depth, $\lambda_{C}$, as a function of an applied magnetic field at several fixed temperatures listed in the legend. The upper panel shows ZFC data, and the bottom panel shows FC results. Also shown is the persistent current density extracted from the conventional magnetization measurements at $T=2\:\text{K}$ using the Bean model \cite{Neha2019}.}
\label{fig:Jc(B)}
\end{figure}

We now examine the magnetic field dependence of the theoretical $j_{c}(H)$ evaluated at several temperatures for both protocols. The upper panel of Fig.\ref{fig:Jc(B)} shows the ZFC data, while the bottom panel shows the FC measurements. As before, solid lines and symbols show $j_{c}(H)$ calculated with temperature-dependent $\xi\left(T\right)$, while dashed lines assume $r_{p}=\xi\left(0\right)=6.4\:\text{nm}$. The two estimates are close and do not change the general picture. For comparison, Fig.\ref{fig:Jc(B)} also shows the persistent current density extracted from conventional magnetization measurements at $T=2\:\text{K}$ estimated from the Bean model \cite{Neha2019}. The overall amplitude is quite comparable, but the actual field dependence is quite different, probably because of the effect of magnetic relaxation during the time window of the measurement. 

The lower panel of Fig.\ref{fig:Jc(B)} presents an unusual result. There is a pronounced peak effect in $j_{c}(H)$ in the FC protocol when vortices oscillate at the bottom of their pinning potential wells in the absence of a vortex-biasing Bean persistent current. This information is inaccessible in conventional magnetization measurements, which are always accompanied by the vortex density gradient, and the measured signal is proportional to this gradient, hence the persistent current density $j_p(H)$. This result implies that the pinning potential changes non-monotonically with increasing magnetic field. Since $j_{c}=\alpha r_{p}/\phi_{0}$, this means that the product $\alpha r_{p}$ is non-monotonic. The Labusch constant does not depend on $r_p$, but depends on the depth of the effective pinning potential, $U_0$. 

In principle, the observed peak effect in a FC protocol can be explained within the collective pinning theory \cite{Blatter1994}, but suggesting a novel mechanism of its formation. As mentioned in the Introduction, there are two explanations of the peak effect: static when there is a nonmonotonic critical current $j_{c}\left(H\right)$, and dynamic in which the peak in the persistent current density is formed as a result of field-dependent magnetic relaxation, which is faster at lower magnetic fields \cite{Blatter1994}.
Our results suggest the existence of the third scenario, which is essentially a combination of the two. The critical current is monotonic in the ZFC state with a vortex density gradient but is non-monotonic in the relaxed FC state without the gradient. This difference is possible for an anharmonic pining potential. However, to arrive at this nonmonotonic peak-effect state, the system needs to relax. The relaxation rate may be a monotonic function of a magnetic field or not.

Of course, a detailed microscopic explanation of the observed ``hidden'' peak effect is likely more complicated. After all, we used a simplified Campbell picture, which may not be quantitatively applicable here. However, we believe that even a simplified discussion captures the key aspects of the results.

\begin{figure}[!tb]
\includegraphics[width=8cm]{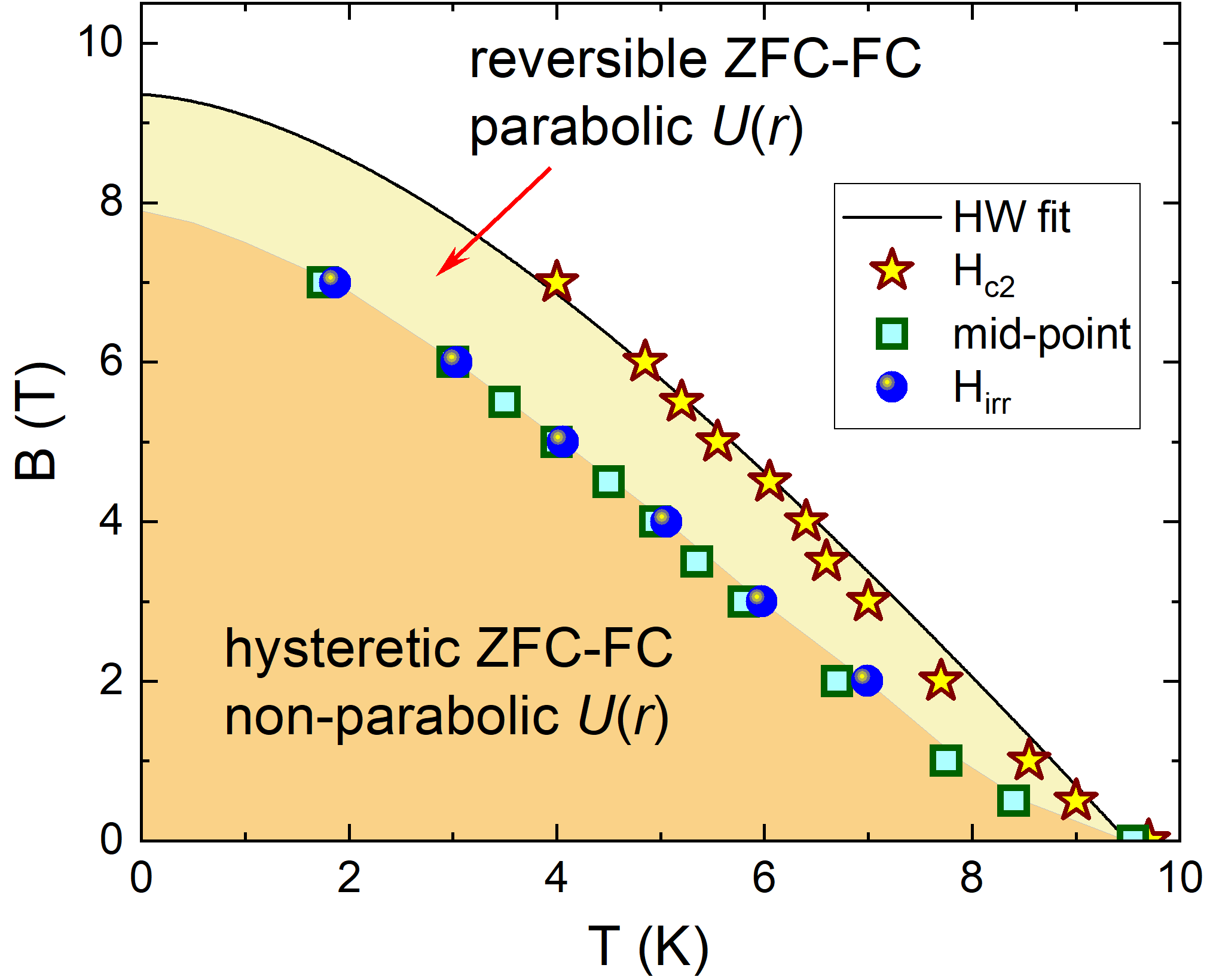} 
\caption{Vortex matter phase diagram constructed from $\lambda_{m}(T,H)$.
The upper critical field, $H_{c2}$$\left(T\right)$, is estimated using
the onset of the diamagnetic transition as a criterion (stars). Also shown
is the irreversibility line below which the ZFC and FC Campbell lengths split
(blue circles). It is practically the same as the midpoint of the transition curves (squares).}
\label{fig:PhaseDiag}
\end{figure}

Finally, we conclude by constructing the $H-T$ vortex phase diagram. Figure \ref{fig:PhaseDiag} shows the upper critical field extracted from the onset of $\lambda_{m}\left(T\right)$ curves shown in Fig.\ref{fig:Lm}. The irreversibility line, obtained as a point where the ZFC and FC curves separate, Fig.\ref{fig:Lm}, is shown by blue symbols. The vertical midpoint line of the $\lambda_{m}\left(T\right)$ dependencies (green squares) follows closely. This phase diagram shows that the effects of anharmonicity diminish at higher temperatures and larger magnetic fields.

\section{Conclusions}

In conclusion, London and Campbell penetration depths were systematically investigated in single crystals of the endohedral gallide cluster superconductor, Mo$_{8}$Ga$_{41}$. The full temperature range superfluid density is consistent with the clean isotropic s-wave weak-coupling BCS theory without any signs of the second gap or strong coupling. The critical current density evaluated from the Campbell length reveals an unusual result. Its field dependence is monotonic in the zero-field cooling (ZFC) process, but exhibits a profound ``hidden'' peak effect in the field-cooling (FC) protocol. It is hidden because there is no vortex density gradient in a FC protocol, whereas conventional measurements of the irreversible state are always accompanied by a vortex density gradient that supports the persistent Bean current. 

We suggest that at least in some compounds, the peak effect appears as a result of magnetic relaxation (long time window of the experiment), but not because this relaxation is magnetic field dependent (which is still possible, though). Instead, the system evolves from an anharmonic regime with monotonic critical current, $j_c(H)$, to a relaxed harmonic regime at smaller persistent current amplitude, $j_p(H) \ll j_c(H)$, where the vortex potential itself is a non-monotonic function of a magnetic field. Therefore, this scenario of peak effect formation involves both static and dynamic aspects. 

\section{Acknowledgments}

We thank V. Geshkenbein for the useful discussions. This work was supported by the US DOE, Office of Science, BES Materials Science and Engineering Division under contract $\#$ DE-AC02-07CH11358. Work at BNL (materials synthesis) was supported by the U.S. Department of Energy, Basic Energy Sciences, Division of Materials Science and Engineering, under Contract No. DE-SC0012704. C.P. acknowledges support from the Shanghai Key Laboratory of Material Frontiers Research in Extreme Environments, China (No. 22dz2260800) and Shanghai Science and Technology Committee, China (No. 22JC1410300).


\begin{thebibliography}{68}%
\makeatletter
\providecommand \@ifxundefined [1]{%
 \@ifx{#1\undefined}
}%
\providecommand \@ifnum [1]{%
 \ifnum #1\expandafter \@firstoftwo
 \else \expandafter \@secondoftwo
 \fi
}%
\providecommand \@ifx [1]{%
 \ifx #1\expandafter \@firstoftwo
 \else \expandafter \@secondoftwo
 \fi
}%
\providecommand \natexlab [1]{#1}%
\providecommand \enquote  [1]{``#1''}%
\providecommand \bibnamefont  [1]{#1}%
\providecommand \bibfnamefont [1]{#1}%
\providecommand \citenamefont [1]{#1}%
\providecommand \href@noop [0]{\@secondoftwo}%
\providecommand \href [0]{\begingroup \@sanitize@url \@href}%
\providecommand \@href[1]{\@@startlink{#1}\@@href}%
\providecommand \@@href[1]{\endgroup#1\@@endlink}%
\providecommand \@sanitize@url [0]{\catcode `\\12\catcode `\$12\catcode
  `\&12\catcode `\#12\catcode `\^12\catcode `\_12\catcode `\%12\relax}%
\providecommand \@@startlink[1]{}%
\providecommand \@@endlink[0]{}%
\providecommand \url  [0]{\begingroup\@sanitize@url \@url }%
\providecommand \@url [1]{\endgroup\@href {#1}{\urlprefix }}%
\providecommand \urlprefix  [0]{URL }%
\providecommand \Eprint [0]{\href }%
\providecommand \doibase [0]{https://doi.org/}%
\providecommand \selectlanguage [0]{\@gobble}%
\providecommand \bibinfo  [0]{\@secondoftwo}%
\providecommand \bibfield  [0]{\@secondoftwo}%
\providecommand \translation [1]{[#1]}%
\providecommand \BibitemOpen [0]{}%
\providecommand \bibitemStop [0]{}%
\providecommand \bibitemNoStop [0]{.\EOS\space}%
\providecommand \EOS [0]{\spacefactor3000\relax}%
\providecommand \BibitemShut  [1]{\csname bibitem#1\endcsname}%
\let\auto@bib@innerbib\@empty
\bibitem [{\citenamefont {Crabtree}\ \emph {et~al.}(2003)\citenamefont
  {Crabtree}, \citenamefont {Kwok}, \citenamefont {Canfield},\ and\
  \citenamefont {Bud’ko}}]{Crabtree2003}%
  \BibitemOpen
  \bibfield  {author} {\bibinfo {author} {\bibfnamefont {G.}~\bibnamefont
  {Crabtree}}, \bibinfo {author} {\bibfnamefont {W.}~\bibnamefont {Kwok}},
  \bibinfo {author} {\bibfnamefont {P.~C.}\ \bibnamefont {Canfield}},\ and\
  \bibinfo {author} {\bibfnamefont {S.~L.}\ \bibnamefont {Bud’ko}},\
  }\bibfield  {title} {\bibinfo {title} {Preface},\ }\href
  {https://doi.org/https://doi.org/10.1016/S0921-4534(02)02554-6} {\bibfield
  {journal} {\bibinfo  {journal} {{Physica C, Special Issue }}\ }\textbf
  {\bibinfo {volume} {385}},\ \bibinfo {pages} {vii} (\bibinfo {year}
  {2003})}\BibitemShut {NoStop}%
\bibitem [{\citenamefont {Kresin}\ \emph {et~al.}(2013)\citenamefont {Kresin},
  \citenamefont {Morawitz},\ and\ \citenamefont {Wolf}}]{Kresin2013}%
  \BibitemOpen
  \bibfield  {author} {\bibinfo {author} {\bibfnamefont {V.}~\bibnamefont
  {Kresin}}, \bibinfo {author} {\bibfnamefont {H.}~\bibnamefont {Morawitz}},\
  and\ \bibinfo {author} {\bibfnamefont {S.}~\bibnamefont {Wolf}},\ }\bibinfo
  {title} {Multigap superconductivity}\ (\bibinfo  {publisher} {{Oxford
  Academic}},\ \bibinfo {year} {2013})\ pp.\ \bibinfo {pages}
  {103--113}\BibitemShut {NoStop}%
\bibitem [{\citenamefont {Suhl}\ \emph {et~al.}(1959)\citenamefont {Suhl},
  \citenamefont {Matthias},\ and\ \citenamefont {Walker}}]{Suhl1959}%
  \BibitemOpen
  \bibfield  {author} {\bibinfo {author} {\bibfnamefont {H.}~\bibnamefont
  {Suhl}}, \bibinfo {author} {\bibfnamefont {B.~T.}\ \bibnamefont {Matthias}},\
  and\ \bibinfo {author} {\bibfnamefont {L.~R.}\ \bibnamefont {Walker}},\
  }\bibfield  {title} {\bibinfo {title} {Bardeen-cooper-schrieffer theory of
  superconductivity in the case of overlapping bands},\ }\href
  {https://doi.org/10.1103/PhysRevLett.3.552} {\bibfield  {journal} {\bibinfo
  {journal} {Phys. Rev. Lett.}\ }\textbf {\bibinfo {volume} {3}},\ \bibinfo
  {pages} {552} (\bibinfo {year} {1959})}\BibitemShut {NoStop}%
\bibitem [{\citenamefont {Moskalenko}(1959)}]{Moskalenko1959}%
  \BibitemOpen
  \bibfield  {author} {\bibinfo {author} {\bibfnamefont {V.~A.}\ \bibnamefont
  {Moskalenko}},\ }\bibfield  {title} {\bibinfo {title} {Superconductivity in
  metals with overlapping energy bands},\ }\href@noop {} {\bibfield  {journal}
  {\bibinfo  {journal} {Fiz. Metal. Metalloved}\ }\textbf {\bibinfo {volume}
  {8}},\ \bibinfo {pages} {2518} (\bibinfo {year} {1959})}\BibitemShut
  {NoStop}%
\bibitem [{\citenamefont {Binnig}\ \emph {et~al.}(1980)\citenamefont {Binnig},
  \citenamefont {Baratoff}, \citenamefont {Hoenig},\ and\ \citenamefont
  {Bednorz}}]{Binnig1980}%
  \BibitemOpen
  \bibfield  {author} {\bibinfo {author} {\bibfnamefont {G.}~\bibnamefont
  {Binnig}}, \bibinfo {author} {\bibfnamefont {A.}~\bibnamefont {Baratoff}},
  \bibinfo {author} {\bibfnamefont {H.~E.}\ \bibnamefont {Hoenig}},\ and\
  \bibinfo {author} {\bibfnamefont {J.~G.}\ \bibnamefont {Bednorz}},\
  }\bibfield  {title} {\bibinfo {title} {{Two-Band Superconductivity in
  Nb-Doped SrTi${\mathrm{O}}_{3}$}},\ }\href
  {https://doi.org/10.1103/PhysRevLett.45.1352} {\bibfield  {journal} {\bibinfo
   {journal} {Phys. Rev. Lett.}\ }\textbf {\bibinfo {volume} {45}},\ \bibinfo
  {pages} {1352} (\bibinfo {year} {1980})}\BibitemShut {NoStop}%
\bibitem [{\citenamefont {Nagamatsu}\ \emph {et~al.}(2001)\citenamefont
  {Nagamatsu}, \citenamefont {Nakagawa}, \citenamefont {Muranaka},
  \citenamefont {Zenitani},\ and\ \citenamefont {Akimitsu}}]{Nagamatsu2001}%
  \BibitemOpen
  \bibfield  {author} {\bibinfo {author} {\bibfnamefont {J.}~\bibnamefont
  {Nagamatsu}}, \bibinfo {author} {\bibfnamefont {N.}~\bibnamefont {Nakagawa}},
  \bibinfo {author} {\bibfnamefont {T.}~\bibnamefont {Muranaka}}, \bibinfo
  {author} {\bibfnamefont {Y.}~\bibnamefont {Zenitani}},\ and\ \bibinfo
  {author} {\bibfnamefont {J.}~\bibnamefont {Akimitsu}},\ }\bibfield  {title}
  {\bibinfo {title} {{S}uperconductivity at 39 {K} in magnesium diboride},\
  }\href@noop {} {\bibfield  {journal} {\bibinfo  {journal} {Nature}\ }\textbf
  {\bibinfo {volume} {410}},\ \bibinfo {pages} {63} (\bibinfo {year}
  {2001})}\BibitemShut {NoStop}%
\bibitem [{\citenamefont {Bezinge}\ \emph {et~al.}(1984)\citenamefont
  {Bezinge}, \citenamefont {Yvon}, \citenamefont {Decroux},\ and\ \citenamefont
  {Muller}}]{sciencedirect_1984}%
  \BibitemOpen
  \bibfield  {author} {\bibinfo {author} {\bibfnamefont {A.}~\bibnamefont
  {Bezinge}}, \bibinfo {author} {\bibfnamefont {K.}~\bibnamefont {Yvon}},
  \bibinfo {author} {\bibfnamefont {M.}~\bibnamefont {Decroux}},\ and\ \bibinfo
  {author} {\bibfnamefont {J.}~\bibnamefont {Muller}},\ }\bibfield  {title}
  {\bibinfo {title} {{On the existence of binary Mo$_{8}$Ga$_{41}$ and its
  properties}},\ }\href
  {https://doi.org/https://doi.org/10.1016/0022-5088(84)90237-6} {\bibfield
  {journal} {\bibinfo  {journal} {J. Less-Common Met.}\ }\textbf {\bibinfo
  {volume} {99}},\ \bibinfo {pages} {L27} (\bibinfo {year} {1984})}\BibitemShut
  {NoStop}%
\bibitem [{\citenamefont {Xie}\ \emph {et~al.}(2015)\citenamefont {Xie},
  \citenamefont {Luo}, \citenamefont {Phelan}, \citenamefont {Klimczuk},
  \citenamefont {Cevallos},\ and\ \citenamefont {Cava}}]{Xie2015}%
  \BibitemOpen
  \bibfield  {author} {\bibinfo {author} {\bibfnamefont {W.}~\bibnamefont
  {Xie}}, \bibinfo {author} {\bibfnamefont {H.}~\bibnamefont {Luo}}, \bibinfo
  {author} {\bibfnamefont {B.~F.}\ \bibnamefont {Phelan}}, \bibinfo {author}
  {\bibfnamefont {T.}~\bibnamefont {Klimczuk}}, \bibinfo {author}
  {\bibfnamefont {F.~A.}\ \bibnamefont {Cevallos}},\ and\ \bibinfo {author}
  {\bibfnamefont {R.~J.}\ \bibnamefont {Cava}},\ }\bibfield  {title} {\bibinfo
  {title} {Endohedral gallide cluster superconductors and superconductivity in
  {ReGa}$_5$},\ }\bibfield  {journal} {\bibinfo  {journal} {Proc. Nat. Acad.
  Sci.}\ }\textbf {\bibinfo {volume} {112}},\ \href
  {https://doi.org/10.1073/pnas.1522191112} {10.1073/pnas.1522191112} (\bibinfo
  {year} {2015})\BibitemShut {NoStop}%
\bibitem [{\citenamefont {Sirohi}\ \emph {et~al.}(2019)\citenamefont {Sirohi},
  \citenamefont {Saha}, \citenamefont {Neha}, \citenamefont {Das},
  \citenamefont {Patnaik}, \citenamefont {Das},\ and\ \citenamefont
  {Sheet}}]{Sirohi2019}%
  \BibitemOpen
  \bibfield  {author} {\bibinfo {author} {\bibfnamefont {A.}~\bibnamefont
  {Sirohi}}, \bibinfo {author} {\bibfnamefont {S.}~\bibnamefont {Saha}},
  \bibinfo {author} {\bibfnamefont {P.}~\bibnamefont {Neha}}, \bibinfo {author}
  {\bibfnamefont {S.}~\bibnamefont {Das}}, \bibinfo {author} {\bibfnamefont
  {S.}~\bibnamefont {Patnaik}}, \bibinfo {author} {\bibfnamefont
  {T.}~\bibnamefont {Das}},\ and\ \bibinfo {author} {\bibfnamefont
  {G.}~\bibnamefont {Sheet}},\ }\bibfield  {title} {\bibinfo {title}
  {{Multiband superconductivity in ${\mathrm{Mo}}_{8}{\mathrm{Ga}}_{41}$ driven
  by a site-selective mechanism}},\ }\href
  {https://doi.org/10.1103/PhysRevB.99.054503} {\bibfield  {journal} {\bibinfo
  {journal} {Phys. Rev. B}\ }\textbf {\bibinfo {volume} {99}},\ \bibinfo
  {pages} {054503} (\bibinfo {year} {2019})}\BibitemShut {NoStop}%
\bibitem [{\citenamefont {Verchenko}\ and\ \citenamefont
  {Shevelkov}(2021)}]{Verchenko2021}%
  \BibitemOpen
  \bibfield  {author} {\bibinfo {author} {\bibfnamefont {V.~Y.}\ \bibnamefont
  {Verchenko}}\ and\ \bibinfo {author} {\bibfnamefont {A.~V.}\ \bibnamefont
  {Shevelkov}},\ }\bibfield  {title} {\bibinfo {title} {Endohedral cluster
  intermetallic superconductors: at the frontier between chemistry and
  physics},\ }\href {https://doi.org/10.1039/D1DT00587A} {\bibfield  {journal}
  {\bibinfo  {journal} {Dalton Trans.}\ }\textbf {\bibinfo {volume} {50}},\
  \bibinfo {pages} {5109} (\bibinfo {year} {2021})}\BibitemShut {NoStop}%
\bibitem [{\citenamefont {Verchenko}\ \emph {et~al.}(2020)\citenamefont
  {Verchenko}, \citenamefont {Zubtsovskii}, \citenamefont {Tsirlin},
  \citenamefont {Wei}, \citenamefont {Roslova}, \citenamefont {Dikarev},\ and\
  \citenamefont {Shevelkov}}]{Verchenko2020}%
  \BibitemOpen
  \bibfield  {author} {\bibinfo {author} {\bibfnamefont {V.~Y.}\ \bibnamefont
  {Verchenko}}, \bibinfo {author} {\bibfnamefont {A.~O.}\ \bibnamefont
  {Zubtsovskii}}, \bibinfo {author} {\bibfnamefont {A.~A.}\ \bibnamefont
  {Tsirlin}}, \bibinfo {author} {\bibfnamefont {Z.}~\bibnamefont {Wei}},
  \bibinfo {author} {\bibfnamefont {M.}~\bibnamefont {Roslova}}, \bibinfo
  {author} {\bibfnamefont {E.~V.}\ \bibnamefont {Dikarev}},\ and\ \bibinfo
  {author} {\bibfnamefont {A.~V.}\ \bibnamefont {Shevelkov}},\ }\bibfield
  {title} {\bibinfo {title} {{Mo$_{8}$Ga$_{41}$ endohedral cluster
  superconductor}},\ }\href
  {https://doi.org/https://doi.org/10.1016/j.jallcom.2020.156400} {\bibfield
  {journal} {\bibinfo  {journal} {J. Alloys Compd.}\ }\textbf {\bibinfo
  {volume} {848}},\ \bibinfo {pages} {156400} (\bibinfo {year}
  {2020})}\BibitemShut {NoStop}%
\bibitem [{\citenamefont {Bardeen}\ \emph
  {et~al.}(1957{\natexlab{a}})\citenamefont {Bardeen}, \citenamefont {Cooper},\
  and\ \citenamefont {Schrieffer}}]{Bardeen1957a}%
  \BibitemOpen
  \bibfield  {author} {\bibinfo {author} {\bibfnamefont {J.}~\bibnamefont
  {Bardeen}}, \bibinfo {author} {\bibfnamefont {L.~N.}\ \bibnamefont
  {Cooper}},\ and\ \bibinfo {author} {\bibfnamefont {J.~R.}\ \bibnamefont
  {Schrieffer}},\ }\bibfield  {title} {\bibinfo {title} {Theory of
  superconductivity},\ }\href {https://doi.org/10.1103/PhysRev.108.1175}
  {\bibfield  {journal} {\bibinfo  {journal} {Phys. Rev.}\ }\textbf {\bibinfo
  {volume} {108}},\ \bibinfo {pages} {1175} (\bibinfo {year}
  {1957}{\natexlab{a}})}\BibitemShut {NoStop}%
\bibitem [{\citenamefont {Bardeen}\ \emph
  {et~al.}(1957{\natexlab{b}})\citenamefont {Bardeen}, \citenamefont {Cooper},\
  and\ \citenamefont {Schrieffer}}]{Bardeen1957}%
  \BibitemOpen
  \bibfield  {author} {\bibinfo {author} {\bibfnamefont {J.}~\bibnamefont
  {Bardeen}}, \bibinfo {author} {\bibfnamefont {L.~N.}\ \bibnamefont
  {Cooper}},\ and\ \bibinfo {author} {\bibfnamefont {J.~R.}\ \bibnamefont
  {Schrieffer}},\ }\bibfield  {title} {\bibinfo {title} {Microscopic theory of
  superconductivity},\ }\href {https://doi.org/10.1103/physrev.106.162}
  {\bibfield  {journal} {\bibinfo  {journal} {Phys. Rev.}\ }\textbf {\bibinfo
  {volume} {106}},\ \bibinfo {pages} {162} (\bibinfo {year}
  {1957}{\natexlab{b}})}\BibitemShut {NoStop}%
\bibitem [{\citenamefont {Hu}\ \emph {et~al.}(2020)\citenamefont {Hu},
  \citenamefont {Graf}, \citenamefont {Liu},\ and\ \citenamefont
  {Petrovic}}]{Petrovic_QO_2020}%
  \BibitemOpen
  \bibfield  {author} {\bibinfo {author} {\bibfnamefont {Z.}~\bibnamefont
  {Hu}}, \bibinfo {author} {\bibfnamefont {D.}~\bibnamefont {Graf}}, \bibinfo
  {author} {\bibfnamefont {Y.}~\bibnamefont {Liu}},\ and\ \bibinfo {author}
  {\bibfnamefont {C.}~\bibnamefont {Petrovic}},\ }\bibfield  {title} {\bibinfo
  {title} {{Three-dimensional Fermi surface and small effective masses in
  Mo$_{8}$Ga$_{41}$}},\ }\href {https://doi.org/10.1063/5.0005177} {\bibfield
  {journal} {\bibinfo  {journal} {Appl. Phys. Lett.}\ }\textbf {\bibinfo
  {volume} {116}},\ \bibinfo {pages} {202601} (\bibinfo {year}
  {2020})}\BibitemShut {NoStop}%
\bibitem [{\citenamefont {Verchenko}\ \emph {et~al.}(2016)\citenamefont
  {Verchenko}, \citenamefont {Tsirlin}, \citenamefont {Zubtsovskiy},\ and\
  \citenamefont {Shevelkov}}]{Verchenko2016}%
  \BibitemOpen
  \bibfield  {author} {\bibinfo {author} {\bibfnamefont {V.~Y.}\ \bibnamefont
  {Verchenko}}, \bibinfo {author} {\bibfnamefont {A.~A.}\ \bibnamefont
  {Tsirlin}}, \bibinfo {author} {\bibfnamefont {A.~O.}\ \bibnamefont
  {Zubtsovskiy}},\ and\ \bibinfo {author} {\bibfnamefont {A.~V.}\ \bibnamefont
  {Shevelkov}},\ }\bibfield  {title} {\bibinfo {title} {{Strong electron-phonon
  coupling in the intermetallic superconductor Mo$_{8}$Ga$_{41}$}},\ }\href
  {https://doi.org/10.1103/PhysRevB.93.064501} {\bibfield  {journal} {\bibinfo
  {journal} {Phys. Rev. B}\ }\textbf {\bibinfo {volume} {93}},\ \bibinfo
  {pages} {064501} (\bibinfo {year} {2016})}\BibitemShut {NoStop}%
\bibitem [{\citenamefont {Verchenko}\ \emph {et~al.}(2017)\citenamefont
  {Verchenko}, \citenamefont {Khasanov}, \citenamefont {Guguchia},
  \citenamefont {Tsirlin},\ and\ \citenamefont {Shevelkov}}]{Verchenko2017}%
  \BibitemOpen
  \bibfield  {author} {\bibinfo {author} {\bibfnamefont {V.~Y.}\ \bibnamefont
  {Verchenko}}, \bibinfo {author} {\bibfnamefont {R.}~\bibnamefont {Khasanov}},
  \bibinfo {author} {\bibfnamefont {Z.}~\bibnamefont {Guguchia}}, \bibinfo
  {author} {\bibfnamefont {A.~A.}\ \bibnamefont {Tsirlin}},\ and\ \bibinfo
  {author} {\bibfnamefont {A.~V.}\ \bibnamefont {Shevelkov}},\ }\bibfield
  {title} {\bibinfo {title} {{Two-gap superconductivity in Mo$_{8}$Ga$_{41}$
  and its evolution upon vanadium substitution}},\ }\href
  {https://doi.org/10.1103/PhysRevB.96.134504} {\bibfield  {journal} {\bibinfo
  {journal} {Phys. Rev. B}\ }\textbf {\bibinfo {volume} {96}},\ \bibinfo
  {pages} {134504} (\bibinfo {year} {2017})}\BibitemShut {NoStop}%
\bibitem [{\citenamefont {Marcin}\ \emph {et~al.}(2021)\citenamefont {Marcin},
  \citenamefont {Pribulov{\'{a}}}, \citenamefont
  {Ka{\v{c}}mar{\v{c}}{\'{\i}}k}, \citenamefont {Medveck{\'{a}}}, \citenamefont
  {Klein}, \citenamefont {Verchenko}, \citenamefont {Cambel}, \citenamefont
  {{\v{S}}olt{\'{y}}s},\ and\ \citenamefont {Samuely}}]{Marcin_2021}%
  \BibitemOpen
  \bibfield  {author} {\bibinfo {author} {\bibfnamefont {M.}~\bibnamefont
  {Marcin}}, \bibinfo {author} {\bibfnamefont {Z.}~\bibnamefont
  {Pribulov{\'{a}}}}, \bibinfo {author} {\bibfnamefont {J.}~\bibnamefont
  {Ka{\v{c}}mar{\v{c}}{\'{\i}}k}}, \bibinfo {author} {\bibfnamefont
  {Z.}~\bibnamefont {Medveck{\'{a}}}}, \bibinfo {author} {\bibfnamefont
  {T.}~\bibnamefont {Klein}}, \bibinfo {author} {\bibfnamefont {V.~Y.}\
  \bibnamefont {Verchenko}}, \bibinfo {author} {\bibfnamefont {V.}~\bibnamefont
  {Cambel}}, \bibinfo {author} {\bibfnamefont {J.}~\bibnamefont
  {{\v{S}}olt{\'{y}}s}},\ and\ \bibinfo {author} {\bibfnamefont
  {P.}~\bibnamefont {Samuely}},\ }\bibfield  {title} {\bibinfo {title} {{One or
  two gaps in Mo$_{8}$Ga$_{41}$ superconductor? Local Hall-probe magnetometry
  study}},\ }\href {https://doi.org/10.1088/1361-6668/abd5f3} {\bibfield
  {journal} {\bibinfo  {journal} {Supercond. Sci. Technol.}\ }\textbf {\bibinfo
  {volume} {34}},\ \bibinfo {pages} {035017} (\bibinfo {year}
  {2021})}\BibitemShut {NoStop}%
\bibitem [{\citenamefont {Marcin}\ \emph {et~al.}(2019)\citenamefont {Marcin},
  \citenamefont {Kačmarčík}, \citenamefont {Pribulová}, \citenamefont
  {Kopčík}, \citenamefont {Szabó}, \citenamefont {Šofranko}, \citenamefont
  {Samuely}, \citenamefont {Vaňo}, \citenamefont {Marcenat}, \citenamefont
  {Verchenko}, \citenamefont {Shevelkov},\ and\ \citenamefont
  {Samuely}}]{Marcin2019}%
  \BibitemOpen
  \bibfield  {author} {\bibinfo {author} {\bibfnamefont {M.}~\bibnamefont
  {Marcin}}, \bibinfo {author} {\bibfnamefont {J.}~\bibnamefont
  {Kačmarčík}}, \bibinfo {author} {\bibfnamefont {Z.}~\bibnamefont
  {Pribulová}}, \bibinfo {author} {\bibfnamefont {M.}~\bibnamefont
  {Kopčík}}, \bibinfo {author} {\bibfnamefont {P.}~\bibnamefont {Szabó}},
  \bibinfo {author} {\bibfnamefont {O.}~\bibnamefont {Šofranko}}, \bibinfo
  {author} {\bibfnamefont {T.}~\bibnamefont {Samuely}}, \bibinfo {author}
  {\bibfnamefont {V.}~\bibnamefont {Vaňo}}, \bibinfo {author} {\bibfnamefont
  {C.}~\bibnamefont {Marcenat}}, \bibinfo {author} {\bibfnamefont {V.~Y.}\
  \bibnamefont {Verchenko}}, \bibinfo {author} {\bibfnamefont {A.~V.}\
  \bibnamefont {Shevelkov}},\ and\ \bibinfo {author} {\bibfnamefont
  {P.}~\bibnamefont {Samuely}},\ }\bibfield  {title} {\bibinfo {title}
  {Single-gap superconductivity in mo$_{8}$ga$_{41}$},\ }\href
  {https://doi.org/10.1038/s41598-019-49846-y} {\bibfield  {journal} {\bibinfo
  {journal} {Scientific Reports}\ }\textbf {\bibinfo {volume} {9}},\ \bibinfo
  {pages} {13552} (\bibinfo {year} {2019})}\BibitemShut {NoStop}%
\bibitem [{\citenamefont {van~der Beek}\ \emph {et~al.}(2002)\citenamefont
  {van~der Beek}, \citenamefont {Konczykowski}, \citenamefont {Abal'oshev},
  \citenamefont {Abal'osheva}, \citenamefont {Gierlowski}, \citenamefont
  {Lewandowski}, \citenamefont {Indenbom},\ and\ \citenamefont
  {Barbanera}}]{vanderBeek2002}%
  \BibitemOpen
  \bibfield  {author} {\bibinfo {author} {\bibfnamefont {C.~J.}\ \bibnamefont
  {van~der Beek}}, \bibinfo {author} {\bibfnamefont {M.}~\bibnamefont
  {Konczykowski}}, \bibinfo {author} {\bibfnamefont {A.}~\bibnamefont
  {Abal'oshev}}, \bibinfo {author} {\bibfnamefont {I.}~\bibnamefont
  {Abal'osheva}}, \bibinfo {author} {\bibfnamefont {P.}~\bibnamefont
  {Gierlowski}}, \bibinfo {author} {\bibfnamefont {S.~J.}\ \bibnamefont
  {Lewandowski}}, \bibinfo {author} {\bibfnamefont {M.~V.}\ \bibnamefont
  {Indenbom}},\ and\ \bibinfo {author} {\bibfnamefont {S.}~\bibnamefont
  {Barbanera}},\ }\bibfield  {title} {\bibinfo {title} {Strong pinning in
  high-temperature superconducting films},\ }\href
  {https://doi.org/10.1103/PhysRevB.66.024523} {\bibfield  {journal} {\bibinfo
  {journal} {Phys. Rev. B}\ }\textbf {\bibinfo {volume} {66}},\ \bibinfo
  {pages} {024523} (\bibinfo {year} {2002})}\BibitemShut {NoStop}%
\bibitem [{\citenamefont {van~der Beek}\ \emph {et~al.}(2012)\citenamefont
  {van~der Beek}, \citenamefont {Konczykowski},\ and\ \citenamefont
  {Prozorov}}]{Beek2012a}%
  \BibitemOpen
  \bibfield  {author} {\bibinfo {author} {\bibfnamefont {C.~J.}\ \bibnamefont
  {van~der Beek}}, \bibinfo {author} {\bibfnamefont {M.}~\bibnamefont
  {Konczykowski}},\ and\ \bibinfo {author} {\bibfnamefont {R.}~\bibnamefont
  {Prozorov}},\ }\bibfield  {title} {\bibinfo {title} {Anisotropy of strong
  pinning in multi-band superconductors},\ }\href
  {https://doi.org/10.1088/0953-2048/25/8/084010} {\bibfield  {journal}
  {\bibinfo  {journal} {Supercond. Sci. Technol.}\ }\textbf {\bibinfo {volume}
  {25}},\ \bibinfo {pages} {084010} (\bibinfo {year} {2012})}\BibitemShut
  {NoStop}%
\bibitem [{\citenamefont {Willa}\ \emph
  {et~al.}(2015{\natexlab{a}})\citenamefont {Willa}, \citenamefont
  {Geshkenbein},\ and\ \citenamefont {Blatter}}]{Willa2015}%
  \BibitemOpen
  \bibfield  {author} {\bibinfo {author} {\bibfnamefont {R.}~\bibnamefont
  {Willa}}, \bibinfo {author} {\bibfnamefont {V.~B.}\ \bibnamefont
  {Geshkenbein}},\ and\ \bibinfo {author} {\bibfnamefont {G.}~\bibnamefont
  {Blatter}},\ }\bibfield  {title} {\bibinfo {title} {{Campbell penetration in
  the critical state of type-II superconductors}},\ }\href
  {https://doi.org/10.1103/PhysRevB.92.134501} {\bibfield  {journal} {\bibinfo
  {journal} {Phys. Rev. B}\ }\textbf {\bibinfo {volume} {92}},\ \bibinfo
  {pages} {134501} (\bibinfo {year} {2015}{\natexlab{a}})}\BibitemShut
  {NoStop}%
\bibitem [{\citenamefont {Willa}\ \emph
  {et~al.}(2015{\natexlab{b}})\citenamefont {Willa}, \citenamefont
  {Geshkenbein}, \citenamefont {Prozorov},\ and\ \citenamefont
  {Blatter}}]{Willa2015a}%
  \BibitemOpen
  \bibfield  {author} {\bibinfo {author} {\bibfnamefont {R.}~\bibnamefont
  {Willa}}, \bibinfo {author} {\bibfnamefont {V.~B.}\ \bibnamefont
  {Geshkenbein}}, \bibinfo {author} {\bibfnamefont {R.}~\bibnamefont
  {Prozorov}},\ and\ \bibinfo {author} {\bibfnamefont {G.}~\bibnamefont
  {Blatter}},\ }\bibfield  {title} {\bibinfo {title} {{Campbell Response in
  Type-II Superconductors under Strong Pinning Conditions}},\ }\href
  {https://doi.org/10.1103/PhysRevLett.115.207001} {\bibfield  {journal}
  {\bibinfo  {journal} {Phys. Rev. Lett.}\ }\textbf {\bibinfo {volume} {115}},\
  \bibinfo {pages} {207001} (\bibinfo {year} {2015}{\natexlab{b}})}\BibitemShut
  {NoStop}%
\bibitem [{\citenamefont {Willa}\ \emph {et~al.}(2016)\citenamefont {Willa},
  \citenamefont {Geshkenbein},\ and\ \citenamefont {Blatter}}]{Willa2016}%
  \BibitemOpen
  \bibfield  {author} {\bibinfo {author} {\bibfnamefont {R.}~\bibnamefont
  {Willa}}, \bibinfo {author} {\bibfnamefont {V.~B.}\ \bibnamefont
  {Geshkenbein}},\ and\ \bibinfo {author} {\bibfnamefont {G.}~\bibnamefont
  {Blatter}},\ }\bibfield  {title} {\bibinfo {title} {{Probing the pinning
  landscape in type-II superconductors via Campbell penetration depth}},\
  }\href {https://doi.org/10.1103/PhysRevB.93.064515} {\bibfield  {journal}
  {\bibinfo  {journal} {Phys. Rev. B}\ }\textbf {\bibinfo {volume} {93}},\
  \bibinfo {pages} {064515} (\bibinfo {year} {2016})}\BibitemShut {NoStop}%
\bibitem [{\citenamefont {Eley}\ \emph {et~al.}(2021)\citenamefont {Eley},
  \citenamefont {Glatz},\ and\ \citenamefont {Willa}}]{Willa2021}%
  \BibitemOpen
  \bibfield  {author} {\bibinfo {author} {\bibfnamefont {S.}~\bibnamefont
  {Eley}}, \bibinfo {author} {\bibfnamefont {A.}~\bibnamefont {Glatz}},\ and\
  \bibinfo {author} {\bibfnamefont {R.}~\bibnamefont {Willa}},\ }\bibfield
  {title} {\bibinfo {title} {Challenges and transformative opportunities in
  superconductor vortex physics},\ }\href {https://doi.org/10.1063/5.0055611}
  {\bibfield  {journal} {\bibinfo  {journal} {J. Appl. Phys.}\ }\textbf
  {\bibinfo {volume} {130}},\ \bibinfo {pages} {050901} (\bibinfo {year}
  {2021})}\BibitemShut {NoStop}%
\bibitem [{\citenamefont {Bean}(1964)}]{Bean1964}%
  \BibitemOpen
  \bibfield  {author} {\bibinfo {author} {\bibfnamefont {C.~P.}\ \bibnamefont
  {Bean}},\ }\bibfield  {title} {\bibinfo {title} {{M}agnetization of
  {H}igh-{F}ield {S}uperconductors},\ }\href
  {http://link.aps.org/abstract/RMP/v36/p31} {\bibfield  {journal} {\bibinfo
  {journal} {Rev. Mod. Phys.}\ }\textbf {\bibinfo {volume} {36}},\ \bibinfo
  {pages} {31} (\bibinfo {year} {1964})}\BibitemShut {NoStop}%
\bibitem [{\citenamefont {Bean}(1962)}]{Bean1962}%
  \BibitemOpen
  \bibfield  {author} {\bibinfo {author} {\bibfnamefont {C.~P.}\ \bibnamefont
  {Bean}},\ }\bibfield  {title} {\bibinfo {title} {{M}agnetization of {H}ard
  {S}uperconductors},\ }\href {http://link.aps.org/abstract/PRL/v8/p250}
  {\bibfield  {journal} {\bibinfo  {journal} {Phys. Rev. Lett.}\ }\textbf
  {\bibinfo {volume} {8}},\ \bibinfo {pages} {250} (\bibinfo {year}
  {1962})}\BibitemShut {NoStop}%
\bibitem [{\citenamefont {Neha}\ \emph {et~al.}(2019)\citenamefont {Neha},
  \citenamefont {Sivaprakash}, \citenamefont {Ishigaki}, \citenamefont
  {Kalaiselvan}, \citenamefont {Manikandan}, \citenamefont {Dhaka},
  \citenamefont {Uwatoko}, \citenamefont {Arumugam},\ and\ \citenamefont
  {Patnaik}}]{Neha2019}%
  \BibitemOpen
  \bibfield  {author} {\bibinfo {author} {\bibfnamefont {P.}~\bibnamefont
  {Neha}}, \bibinfo {author} {\bibfnamefont {P.}~\bibnamefont {Sivaprakash}},
  \bibinfo {author} {\bibfnamefont {K.}~\bibnamefont {Ishigaki}}, \bibinfo
  {author} {\bibfnamefont {G.}~\bibnamefont {Kalaiselvan}}, \bibinfo {author}
  {\bibfnamefont {K.}~\bibnamefont {Manikandan}}, \bibinfo {author}
  {\bibfnamefont {R.~S.}\ \bibnamefont {Dhaka}}, \bibinfo {author}
  {\bibfnamefont {Y.}~\bibnamefont {Uwatoko}}, \bibinfo {author} {\bibfnamefont
  {S.}~\bibnamefont {Arumugam}},\ and\ \bibinfo {author} {\bibfnamefont
  {S.}~\bibnamefont {Patnaik}},\ }\bibfield  {title} {\bibinfo {title} {Nuanced
  superconductivity in endohedral gallide {Mo}$_8${Ga}$_{41}$},\ }\href
  {https://doi.org/10.1088/2053-1591/aae5b5} {\bibfield  {journal} {\bibinfo
  {journal} {Mater. Res. Express}\ }\textbf {\bibinfo {volume} {6}},\ \bibinfo
  {pages} {016002} (\bibinfo {year} {2019})}\BibitemShut {NoStop}%
\bibitem [{\citenamefont {Marcin}\ \emph {et~al.}(2020)\citenamefont {Marcin},
  \citenamefont {Pribulov\'{a}}, \citenamefont {Kacmarcik}, \citenamefont
  {Verchenko}, \citenamefont {Shevelkov}, \citenamefont {Cambel}, \citenamefont
  {\v{S}olt\'{y}s},\ and\ \citenamefont {Samuely}}]{Marcin2020}%
  \BibitemOpen
  \bibfield  {author} {\bibinfo {author} {\bibfnamefont {M.}~\bibnamefont
  {Marcin}}, \bibinfo {author} {\bibfnamefont {Z.}~\bibnamefont
  {Pribulov\'{a}}}, \bibinfo {author} {\bibfnamefont {J.}~\bibnamefont
  {Kacmarcik}}, \bibinfo {author} {\bibfnamefont {V.}~\bibnamefont
  {Verchenko}}, \bibinfo {author} {\bibfnamefont {A.}~\bibnamefont
  {Shevelkov}}, \bibinfo {author} {\bibfnamefont {V.}~\bibnamefont {Cambel}},
  \bibinfo {author} {\bibfnamefont {J.}~\bibnamefont {\v{S}olt\'{y}s}},\ and\
  \bibinfo {author} {\bibfnamefont {P.}~\bibnamefont {Samuely}},\ }\bibfield
  {title} {\bibinfo {title} {{Local Magnetometry of Superconducting
  Mo$_{8}$Ga$_{41}$ and Mo$_{7}$VGa$_{41}$ : Vortex Pinning Study}},\ }\href
  {https://doi.org/10.12693/APhysPolA.137.794} {\bibfield  {journal} {\bibinfo
  {journal} {Acta Physica Polonica A}\ }\textbf {\bibinfo {volume} {137}},\
  \bibinfo {pages} {794} (\bibinfo {year} {2020})}\BibitemShut {NoStop}%
\bibitem [{\citenamefont {DeSorbo}(1964)}]{DeSorbo1964}%
  \BibitemOpen
  \bibfield  {author} {\bibinfo {author} {\bibfnamefont {W.}~\bibnamefont
  {DeSorbo}},\ }\bibfield  {title} {\bibinfo {title} {{The Peak Effect in
  Substitutional and Interstitial Solid Solutions of High-Field
  Superconductors}},\ }\href {http://link.aps.org/abstract/RMP/v36/p90}
  {\bibfield  {journal} {\bibinfo  {journal} {Rev. Mod. Phys.}\ }\textbf
  {\bibinfo {volume} {36}},\ \bibinfo {pages} {90} (\bibinfo {year}
  {1964})}\BibitemShut {NoStop}%
\bibitem [{\citenamefont {Larkin}\ and\ \citenamefont
  {Ovchinnikov}(1979)}]{Larkin1979}%
  \BibitemOpen
  \bibfield  {author} {\bibinfo {author} {\bibfnamefont {A.~I.}\ \bibnamefont
  {Larkin}}\ and\ \bibinfo {author} {\bibfnamefont {Y.~N.}\ \bibnamefont
  {Ovchinnikov}},\ }\bibfield  {title} {\bibinfo {title} {{Pinning in type-II
  superconductors}},\ }\href {https://doi.org/10.1007/bf00117160} {\bibfield
  {journal} {\bibinfo  {journal} {J. Low Temp. Phys.}\ }\textbf {\bibinfo
  {volume} {34}},\ \bibinfo {pages} {409} (\bibinfo {year} {1979})}\BibitemShut
  {NoStop}%
\bibitem [{\citenamefont {Daeumling}\ \emph {et~al.}(1990)\citenamefont
  {Daeumling}, \citenamefont {Seuntjens},\ and\ \citenamefont
  {Larbalestier}}]{Daeumling1990}%
  \BibitemOpen
  \bibfield  {author} {\bibinfo {author} {\bibfnamefont {M.}~\bibnamefont
  {Daeumling}}, \bibinfo {author} {\bibfnamefont {J.~M.}\ \bibnamefont
  {Seuntjens}},\ and\ \bibinfo {author} {\bibfnamefont {D.~C.}\ \bibnamefont
  {Larbalestier}},\ }\bibfield  {title} {\bibinfo {title} {{Oxygen-defect flux
  pinning, anomalous magnetization and intra-grain granularity in
  YBa$_2$Cu$_3$0$_{7-\delta}$}},\ }\href {https://doi.org/10.1038/346332a0}
  {\bibfield  {journal} {\bibinfo  {journal} {Nature}\ }\textbf {\bibinfo
  {volume} {346}},\ \bibinfo {pages} {332} (\bibinfo {year}
  {1990})}\BibitemShut {NoStop}%
\bibitem [{\citenamefont {Krusin-Elbaum}\ \emph {et~al.}(1992)\citenamefont
  {Krusin-Elbaum}, \citenamefont {Civale}, \citenamefont {Vinokur},\ and\
  \citenamefont {Holtzberg}}]{KrusinElbaum1992}%
  \BibitemOpen
  \bibfield  {author} {\bibinfo {author} {\bibfnamefont {L.}~\bibnamefont
  {Krusin-Elbaum}}, \bibinfo {author} {\bibfnamefont {L.}~\bibnamefont
  {Civale}}, \bibinfo {author} {\bibfnamefont {V.~M.}\ \bibnamefont
  {Vinokur}},\ and\ \bibinfo {author} {\bibfnamefont {F.}~\bibnamefont
  {Holtzberg}},\ }\bibfield  {title} {\bibinfo {title} {{```Phase diagram''' of
  the vortex-solid phase in Y-Ba-Cu-O crystals: A crossover from single-vortex
  (1D) to collective (3D) pinning regimes}},\ }\href
  {https://doi.org/10.1103/physrevlett.69.2280} {\bibfield  {journal} {\bibinfo
   {journal} {Phys. Rev. Lett.}\ }\textbf {\bibinfo {volume} {69}},\ \bibinfo
  {pages} {2280} (\bibinfo {year} {1992})}\BibitemShut {NoStop}%
\bibitem [{\citenamefont {Klein}\ \emph {et~al.}(1994)\citenamefont {Klein},
  \citenamefont {Yacoby}, \citenamefont {Yeshurun}, \citenamefont {Erb},
  \citenamefont {M\"{u}ller-Vogt}, \citenamefont {Breit},\ and\ \citenamefont
  {W\"{u}hl}}]{Klein1994}%
  \BibitemOpen
  \bibfield  {author} {\bibinfo {author} {\bibfnamefont {L.}~\bibnamefont
  {Klein}}, \bibinfo {author} {\bibfnamefont {E.~R.}\ \bibnamefont {Yacoby}},
  \bibinfo {author} {\bibfnamefont {Y.}~\bibnamefont {Yeshurun}}, \bibinfo
  {author} {\bibfnamefont {A.}~\bibnamefont {Erb}}, \bibinfo {author}
  {\bibfnamefont {G.}~\bibnamefont {M\"{u}ller-Vogt}}, \bibinfo {author}
  {\bibfnamefont {V.}~\bibnamefont {Breit}},\ and\ \bibinfo {author}
  {\bibfnamefont {H.}~\bibnamefont {W\"{u}hl}},\ }\bibfield  {title} {\bibinfo
  {title} {{Peak effect and scaling of irreversible properties in untwinned
  Y-Ba-Cu-O crystals}},\ }\href {https://doi.org/10.1103/physrevb.49.4403}
  {\bibfield  {journal} {\bibinfo  {journal} {Phys. Rev. B}\ }\textbf {\bibinfo
  {volume} {49}},\ \bibinfo {pages} {4403} (\bibinfo {year}
  {1994})}\BibitemShut {NoStop}%
\bibitem [{\citenamefont {Blatter}\ \emph {et~al.}(1994)\citenamefont
  {Blatter}, \citenamefont {Feigel'man}, \citenamefont {Geshkenbein},
  \citenamefont {Larkin},\ and\ \citenamefont {Vinokur}}]{Blatter1994}%
  \BibitemOpen
  \bibfield  {author} {\bibinfo {author} {\bibfnamefont {G.}~\bibnamefont
  {Blatter}}, \bibinfo {author} {\bibfnamefont {M.~V.}\ \bibnamefont
  {Feigel'man}}, \bibinfo {author} {\bibfnamefont {V.~B.}\ \bibnamefont
  {Geshkenbein}}, \bibinfo {author} {\bibfnamefont {A.~I.}\ \bibnamefont
  {Larkin}},\ and\ \bibinfo {author} {\bibfnamefont {V.~M.}\ \bibnamefont
  {Vinokur}},\ }\bibfield  {title} {\bibinfo {title} {{V}ortices in
  high-temperature superconductors},\ }\href
  {https://doi.org/10.1103/RevModPhys.66.1125} {\bibfield  {journal} {\bibinfo
  {journal} {Rev. Mod. Phys.}\ }\textbf {\bibinfo {volume} {66}},\ \bibinfo
  {pages} {1125} (\bibinfo {year} {1994})}\BibitemShut {NoStop}%
\bibitem [{\citenamefont {Brandt}(1995)}]{Brandt_1995}%
  \BibitemOpen
  \bibfield  {author} {\bibinfo {author} {\bibfnamefont {E.~H.}\ \bibnamefont
  {Brandt}},\ }\bibfield  {title} {\bibinfo {title} {The flux-line lattice in
  superconductors},\ }\href {https://doi.org/10.1088/0034-4885/58/11/003}
  {\bibfield  {journal} {\bibinfo  {journal} {Rep. Prog. Phys.}\ }\textbf
  {\bibinfo {volume} {58}},\ \bibinfo {pages} {1465} (\bibinfo {year}
  {1995})}\BibitemShut {NoStop}%
\bibitem [{\citenamefont {Tang}\ \emph {et~al.}(1996)\citenamefont {Tang},
  \citenamefont {Ling}, \citenamefont {Bhattacharya},\ and\ \citenamefont
  {Chaikin}}]{Tang1996}%
  \BibitemOpen
  \bibfield  {author} {\bibinfo {author} {\bibfnamefont {C.}~\bibnamefont
  {Tang}}, \bibinfo {author} {\bibfnamefont {X.}~\bibnamefont {Ling}}, \bibinfo
  {author} {\bibfnamefont {S.}~\bibnamefont {Bhattacharya}},\ and\ \bibinfo
  {author} {\bibfnamefont {P.~M.}\ \bibnamefont {Chaikin}},\ }\bibfield
  {title} {\bibinfo {title} {{Peak effect in superconductors: melting of Larkin
  domains}},\ }\href {https://doi.org/10.1209/epl/i1996-00157-x} {\bibfield
  {journal} {\bibinfo  {journal} {Europhysics Letters (EPL)}\ }\textbf
  {\bibinfo {volume} {35}},\ \bibinfo {pages} {597} (\bibinfo {year}
  {1996})}\BibitemShut {NoStop}%
\bibitem [{\citenamefont {Giller}\ \emph {et~al.}(1997)\citenamefont {Giller},
  \citenamefont {Shaulov}, \citenamefont {Prozorov}, \citenamefont {Abulafia},
  \citenamefont {Wolfus}, \citenamefont {Burlachkov}, \citenamefont {Yeshurun},
  \citenamefont {Zeldov}, \citenamefont {Vinokur}, \citenamefont {Peng},\ and\
  \citenamefont {Greene}}]{Giller1997}%
  \BibitemOpen
  \bibfield  {author} {\bibinfo {author} {\bibfnamefont {D.}~\bibnamefont
  {Giller}}, \bibinfo {author} {\bibfnamefont {A.}~\bibnamefont {Shaulov}},
  \bibinfo {author} {\bibfnamefont {R.}~\bibnamefont {Prozorov}}, \bibinfo
  {author} {\bibfnamefont {Y.}~\bibnamefont {Abulafia}}, \bibinfo {author}
  {\bibfnamefont {Y.}~\bibnamefont {Wolfus}}, \bibinfo {author} {\bibfnamefont
  {L.}~\bibnamefont {Burlachkov}}, \bibinfo {author} {\bibfnamefont
  {Y.}~\bibnamefont {Yeshurun}}, \bibinfo {author} {\bibfnamefont
  {E.}~\bibnamefont {Zeldov}}, \bibinfo {author} {\bibfnamefont {V.~M.}\
  \bibnamefont {Vinokur}}, \bibinfo {author} {\bibfnamefont {J.~L.}\
  \bibnamefont {Peng}},\ and\ \bibinfo {author} {\bibfnamefont {R.~L.}\
  \bibnamefont {Greene}},\ }\bibfield  {title} {\bibinfo {title}
  {{Disorder-induced transition to entangled vortex solid in Nd-Ce-Cu-O
  crystal}},\ }\href@noop {} {\bibfield  {journal} {\bibinfo  {journal} {Phys.
  Rev. Lett.}\ }\textbf {\bibinfo {volume} {79}},\ \bibinfo {pages} {2542}
  (\bibinfo {year} {1997})}\BibitemShut {NoStop}%
\bibitem [{\citenamefont {Banerjee}\ \emph {et~al.}(2000)\citenamefont
  {Banerjee}, \citenamefont {Ramakrishnan}, \citenamefont {Grover},
  \citenamefont {Ravikumar}, \citenamefont {Mishra}, \citenamefont {Sahni},
  \citenamefont {Tomy}, \citenamefont {Balakrishnan}, \citenamefont {Paul},
  \citenamefont {Gammel}, \citenamefont {Bishop}, \citenamefont {Bucher},
  \citenamefont {Higgins},\ and\ \citenamefont {Bhattacharya}}]{Banerjee2000}%
  \BibitemOpen
  \bibfield  {author} {\bibinfo {author} {\bibfnamefont {S.~S.}\ \bibnamefont
  {Banerjee}}, \bibinfo {author} {\bibfnamefont {S.}~\bibnamefont
  {Ramakrishnan}}, \bibinfo {author} {\bibfnamefont {A.~K.}\ \bibnamefont
  {Grover}}, \bibinfo {author} {\bibfnamefont {G.}~\bibnamefont {Ravikumar}},
  \bibinfo {author} {\bibfnamefont {P.~K.}\ \bibnamefont {Mishra}}, \bibinfo
  {author} {\bibfnamefont {V.~C.}\ \bibnamefont {Sahni}}, \bibinfo {author}
  {\bibfnamefont {C.~V.}\ \bibnamefont {Tomy}}, \bibinfo {author}
  {\bibfnamefont {G.}~\bibnamefont {Balakrishnan}}, \bibinfo {author}
  {\bibfnamefont {D.~M.}\ \bibnamefont {Paul}}, \bibinfo {author}
  {\bibfnamefont {P.~L.}\ \bibnamefont {Gammel}}, \bibinfo {author}
  {\bibfnamefont {D.~J.}\ \bibnamefont {Bishop}}, \bibinfo {author}
  {\bibfnamefont {E.}~\bibnamefont {Bucher}}, \bibinfo {author} {\bibfnamefont
  {M.~J.}\ \bibnamefont {Higgins}},\ and\ \bibinfo {author} {\bibfnamefont
  {S.}~\bibnamefont {Bhattacharya}},\ }\bibfield  {title} {\bibinfo {title}
  {{Peak effect, plateau effect, and fishtail anomaly: The reentrant
  amorphization of vortex matter in 2H-NbSe2}},\ }\href@noop {} {\bibfield
  {journal} {\bibinfo  {journal} {Phys. Rev. B}\ }\textbf {\bibinfo {volume}
  {62}},\ \bibinfo {pages} {11838} (\bibinfo {year} {2000})}\BibitemShut
  {NoStop}%
\bibitem [{\citenamefont {Mikitik}\ and\ \citenamefont
  {Brandt}(2001)}]{Mikitik2001}%
  \BibitemOpen
  \bibfield  {author} {\bibinfo {author} {\bibfnamefont {G.~P.}\ \bibnamefont
  {Mikitik}}\ and\ \bibinfo {author} {\bibfnamefont {E.~H.}\ \bibnamefont
  {Brandt}},\ }\bibfield  {title} {\bibinfo {title} {{Peak effect,
  vortex-lattice melting line, and order-disorder transition in conventional
  and high-T-c superconductors}},\ }\href@noop {} {\bibfield  {journal}
  {\bibinfo  {journal} {Phys. Rev. B}\ }\textbf {\bibinfo {volume} {64}},\
  \bibinfo {pages} {184514} (\bibinfo {year} {2001})}\BibitemShut {NoStop}%
\bibitem [{\citenamefont {Prozorov}\ \emph {et~al.}(2008)\citenamefont
  {Prozorov}, \citenamefont {Ni}, \citenamefont {Tanatar}, \citenamefont
  {Kogan}, \citenamefont {Gordon}, \citenamefont {Martin}, \citenamefont
  {Blomberg}, \citenamefont {Prommapan}, \citenamefont {Yan}, \citenamefont
  {Bud'ko},\ and\ \citenamefont {Canfield}}]{Prozorov2008}%
  \BibitemOpen
  \bibfield  {author} {\bibinfo {author} {\bibfnamefont {R.}~\bibnamefont
  {Prozorov}}, \bibinfo {author} {\bibfnamefont {N.}~\bibnamefont {Ni}},
  \bibinfo {author} {\bibfnamefont {M.~A.}\ \bibnamefont {Tanatar}}, \bibinfo
  {author} {\bibfnamefont {V.~G.}\ \bibnamefont {Kogan}}, \bibinfo {author}
  {\bibfnamefont {R.~T.}\ \bibnamefont {Gordon}}, \bibinfo {author}
  {\bibfnamefont {C.}~\bibnamefont {Martin}}, \bibinfo {author} {\bibfnamefont
  {E.~C.}\ \bibnamefont {Blomberg}}, \bibinfo {author} {\bibfnamefont
  {P.}~\bibnamefont {Prommapan}}, \bibinfo {author} {\bibfnamefont {J.~Q.}\
  \bibnamefont {Yan}}, \bibinfo {author} {\bibfnamefont {S.~L.}\ \bibnamefont
  {Bud'ko}},\ and\ \bibinfo {author} {\bibfnamefont {P.~C.}\ \bibnamefont
  {Canfield}},\ }\bibfield  {title} {\bibinfo {title} {{Vortex phase diagram of
  Ba(Fe$_{0.93}$Co$_{0.07}$)$_2$As$_2$ single crystals}},\ }\href {<Go to
  ISI>://WOS:000262245200074 http://prb.aps.org/pdf/PRB/v78/i22/e224506}
  {\bibfield  {journal} {\bibinfo  {journal} {Phys. Rev. B}\ }\textbf {\bibinfo
  {volume} {78}},\ \bibinfo {pages} {224506} (\bibinfo {year}
  {2008})}\BibitemShut {NoStop}%
\bibitem [{\citenamefont {Yeshurun}\ \emph {et~al.}(1996)\citenamefont
  {Yeshurun}, \citenamefont {Malozemoff},\ and\ \citenamefont
  {Shaulov}}]{Yeshurun1996}%
  \BibitemOpen
  \bibfield  {author} {\bibinfo {author} {\bibfnamefont {Y.}~\bibnamefont
  {Yeshurun}}, \bibinfo {author} {\bibfnamefont {A.~P.}\ \bibnamefont
  {Malozemoff}},\ and\ \bibinfo {author} {\bibfnamefont {A.}~\bibnamefont
  {Shaulov}},\ }\bibfield  {title} {\bibinfo {title} {{Magnetic relaxation in
  high-temperature superconductors}},\ }\href
  {https://doi.org/10.1103/revmodphys.68.911} {\bibfield  {journal} {\bibinfo
  {journal} {Rev. Mod. Phys.}\ }\textbf {\bibinfo {volume} {68}},\ \bibinfo
  {pages} {911} (\bibinfo {year} {1996})}\BibitemShut {NoStop}%
\bibitem [{\citenamefont {Pippard}(1969)}]{Pippard1969}%
  \BibitemOpen
  \bibfield  {author} {\bibinfo {author} {\bibfnamefont {A.~B.}\ \bibnamefont
  {Pippard}},\ }\bibfield  {title} {\bibinfo {title} {{A possible mechanism for
  the peak effect in type II superconductors}},\ }\href
  {https://doi.org/10.1080/14786436908217779} {\bibfield  {journal} {\bibinfo
  {journal} {Phil. Mag.}\ }\textbf {\bibinfo {volume} {19}},\ \bibinfo {pages}
  {217} (\bibinfo {year} {1969})}\BibitemShut {NoStop}%
\bibitem [{\citenamefont {Konczykowski}\ \emph {et~al.}(2000)\citenamefont
  {Konczykowski}, \citenamefont {Colson}, \citenamefont {{van der Beek}},
  \citenamefont {Indenbom}, \citenamefont {Kes},\ and\ \citenamefont
  {Zeldov}}]{Konczykowski2000}%
  \BibitemOpen
  \bibfield  {author} {\bibinfo {author} {\bibfnamefont {M.}~\bibnamefont
  {Konczykowski}}, \bibinfo {author} {\bibfnamefont {S.}~\bibnamefont
  {Colson}}, \bibinfo {author} {\bibfnamefont {C.~J.}\ \bibnamefont {{van der
  Beek}}}, \bibinfo {author} {\bibfnamefont {M.~V.}\ \bibnamefont {Indenbom}},
  \bibinfo {author} {\bibfnamefont {P.~H.}\ \bibnamefont {Kes}},\ and\ \bibinfo
  {author} {\bibfnamefont {E.}~\bibnamefont {Zeldov}},\ }\bibfield  {title}
  {\bibinfo {title} {{Magnetic relaxation in the vicinity of second
  magnetization peak in BSCCO crystals}},\ }\href
  {https://doi.org/https://doi.org/10.1016/S0921-4534(99)00675-9} {\bibfield
  {journal} {\bibinfo  {journal} {Physica C}\ }\textbf {\bibinfo {volume}
  {332}},\ \bibinfo {pages} {219} (\bibinfo {year} {2000})}\BibitemShut
  {NoStop}%
\bibitem [{\citenamefont {Abulafia}\ \emph {et~al.}(1996)\citenamefont
  {Abulafia}, \citenamefont {Shaulov}, \citenamefont {Wolfus}, \citenamefont
  {Prozorov}, \citenamefont {Burlachkov}, \citenamefont {Yeshurun},
  \citenamefont {Majer}, \citenamefont {Zeldov}, \citenamefont {Wuhl},
  \citenamefont {Geshkenbein},\ and\ \citenamefont {Vinokur}}]{Abulafia1996}%
  \BibitemOpen
  \bibfield  {author} {\bibinfo {author} {\bibfnamefont {Y.}~\bibnamefont
  {Abulafia}}, \bibinfo {author} {\bibfnamefont {A.}~\bibnamefont {Shaulov}},
  \bibinfo {author} {\bibfnamefont {Y.}~\bibnamefont {Wolfus}}, \bibinfo
  {author} {\bibfnamefont {R.}~\bibnamefont {Prozorov}}, \bibinfo {author}
  {\bibfnamefont {L.}~\bibnamefont {Burlachkov}}, \bibinfo {author}
  {\bibfnamefont {Y.}~\bibnamefont {Yeshurun}}, \bibinfo {author}
  {\bibfnamefont {D.}~\bibnamefont {Majer}}, \bibinfo {author} {\bibfnamefont
  {E.}~\bibnamefont {Zeldov}}, \bibinfo {author} {\bibfnamefont
  {H.}~\bibnamefont {Wuhl}}, \bibinfo {author} {\bibfnamefont {V.~B.}\
  \bibnamefont {Geshkenbein}},\ and\ \bibinfo {author} {\bibfnamefont {V.~M.}\
  \bibnamefont {Vinokur}},\ }\bibfield  {title} {\bibinfo {title} {{Plastic
  vortex creep in YBa$_2$Cu$_3$O$_{7-x}$ crystals}},\ }\href@noop {} {\bibfield
   {journal} {\bibinfo  {journal} {Phys. Rev. Lett.}\ }\textbf {\bibinfo
  {volume} {77}},\ \bibinfo {pages} {1596} (\bibinfo {year}
  {1996})}\BibitemShut {NoStop}%
\bibitem [{\citenamefont {Campbell}(1969)}]{Campbell_1969}%
  \BibitemOpen
  \bibfield  {author} {\bibinfo {author} {\bibfnamefont {A.~M.}\ \bibnamefont
  {Campbell}},\ }\bibfield  {title} {\bibinfo {title} {The response of pinned
  flux vortices to low-frequency fields},\ }\href
  {https://doi.org/10.1088/0022-3719/2/8/318} {\bibfield  {journal} {\bibinfo
  {journal} {J. Phys. C: Solid State Phys.}\ }\textbf {\bibinfo {volume} {2}},\
  \bibinfo {pages} {1492} (\bibinfo {year} {1969})}\BibitemShut {NoStop}%
\bibitem [{\citenamefont {Campbell}(1971)}]{Campbell_1971}%
  \BibitemOpen
  \bibfield  {author} {\bibinfo {author} {\bibfnamefont {A.~M.}\ \bibnamefont
  {Campbell}},\ }\bibfield  {title} {\bibinfo {title} {The interaction distance
  between flux lines and pinning centres},\ }\href
  {https://doi.org/10.1088/0022-3719/4/18/023} {\bibfield  {journal} {\bibinfo
  {journal} {J. Phys. C: Solid State Phys.}\ }\textbf {\bibinfo {volume} {4}},\
  \bibinfo {pages} {3186} (\bibinfo {year} {1971})}\BibitemShut {NoStop}%
\bibitem [{\citenamefont {Koshelev}\ and\ \citenamefont
  {Vinokur}(1991)}]{Koshelev1991}%
  \BibitemOpen
  \bibfield  {author} {\bibinfo {author} {\bibfnamefont {A.~E.}\ \bibnamefont
  {Koshelev}}\ and\ \bibinfo {author} {\bibfnamefont {V.~M.}\ \bibnamefont
  {Vinokur}},\ }\bibfield  {title} {\bibinfo {title} {{Frequency response of
  pinned vortex lattice}},\ }\href@noop {} {\bibfield  {journal} {\bibinfo
  {journal} {{Physica C}}\ }\textbf {\bibinfo {volume} {173}},\ \bibinfo
  {pages} {465} (\bibinfo {year} {1991})}\BibitemShut {NoStop}%
\bibitem [{\citenamefont {Prommapan}\ \emph {et~al.}(2011)\citenamefont
  {Prommapan}, \citenamefont {Tanatar}, \citenamefont {Lee}, \citenamefont
  {Khim}, \citenamefont {Kim},\ and\ \citenamefont
  {Prozorov}}]{Propmann_PRB_2011}%
  \BibitemOpen
  \bibfield  {author} {\bibinfo {author} {\bibfnamefont {P.}~\bibnamefont
  {Prommapan}}, \bibinfo {author} {\bibfnamefont {M.~A.}\ \bibnamefont
  {Tanatar}}, \bibinfo {author} {\bibfnamefont {B.}~\bibnamefont {Lee}},
  \bibinfo {author} {\bibfnamefont {S.}~\bibnamefont {Khim}}, \bibinfo {author}
  {\bibfnamefont {K.~H.}\ \bibnamefont {Kim}},\ and\ \bibinfo {author}
  {\bibfnamefont {R.}~\bibnamefont {Prozorov}},\ }\bibfield  {title} {\bibinfo
  {title} {{Magnetic-field-dependent pinning potential in LiFeAs superconductor
  from its Campbell penetration depth}},\ }\href
  {https://doi.org/10.1103/PhysRevB.84.060509} {\bibfield  {journal} {\bibinfo
  {journal} {Phys. Rev. B}\ }\textbf {\bibinfo {volume} {84}},\ \bibinfo
  {pages} {060509} (\bibinfo {year} {2011})}\BibitemShut {NoStop}%
\bibitem [{\citenamefont {Nusran}\ \emph {et~al.}(2018)\citenamefont {Nusran},
  \citenamefont {Joshi}, \citenamefont {Cho}, \citenamefont {Tanatar},
  \citenamefont {Meier}, \citenamefont {Bud'ko}, \citenamefont {Canfield},
  \citenamefont {Liu}, \citenamefont {Lograsso},\ and\ \citenamefont
  {Prozorov}}]{Nusran2018}%
  \BibitemOpen
  \bibfield  {author} {\bibinfo {author} {\bibfnamefont {N.~M.}\ \bibnamefont
  {Nusran}}, \bibinfo {author} {\bibfnamefont {K.~R.}\ \bibnamefont {Joshi}},
  \bibinfo {author} {\bibfnamefont {K.}~\bibnamefont {Cho}}, \bibinfo {author}
  {\bibfnamefont {M.~A.}\ \bibnamefont {Tanatar}}, \bibinfo {author}
  {\bibfnamefont {W.~R.}\ \bibnamefont {Meier}}, \bibinfo {author}
  {\bibfnamefont {S.~L.}\ \bibnamefont {Bud'ko}}, \bibinfo {author}
  {\bibfnamefont {P.~C.}\ \bibnamefont {Canfield}}, \bibinfo {author}
  {\bibfnamefont {Y.}~\bibnamefont {Liu}}, \bibinfo {author} {\bibfnamefont
  {T.~A.}\ \bibnamefont {Lograsso}},\ and\ \bibinfo {author} {\bibfnamefont
  {R.}~\bibnamefont {Prozorov}},\ }\bibfield  {title} {\bibinfo {title}
  {Spatially-resolved study of the meissner effect in superconductors using
  {N}{V}-centers-in-diamond optical magnetometry},\ }\href
  {http://stacks.iop.org/1367-2630/20/i=4/a=043010} {\bibfield  {journal}
  {\bibinfo  {journal} {New J. Phys.}\ }\textbf {\bibinfo {volume} {20}},\
  \bibinfo {pages} {043010} (\bibinfo {year} {2018})}\BibitemShut {NoStop}%
\bibitem [{\citenamefont {Joshi}\ \emph {et~al.}(2019)\citenamefont {Joshi},
  \citenamefont {Nusran}, \citenamefont {Tanatar}, \citenamefont {Cho},
  \citenamefont {Meier}, \citenamefont {Bud'ko}, \citenamefont {Canfield},\
  and\ \citenamefont {Prozorov}}]{Joshi2019}%
  \BibitemOpen
  \bibfield  {author} {\bibinfo {author} {\bibfnamefont {K.}~\bibnamefont
  {Joshi}}, \bibinfo {author} {\bibfnamefont {N.}~\bibnamefont {Nusran}},
  \bibinfo {author} {\bibfnamefont {M.}~\bibnamefont {Tanatar}}, \bibinfo
  {author} {\bibfnamefont {K.}~\bibnamefont {Cho}}, \bibinfo {author}
  {\bibfnamefont {W.}~\bibnamefont {Meier}}, \bibinfo {author} {\bibfnamefont
  {S.}~\bibnamefont {Bud'ko}}, \bibinfo {author} {\bibfnamefont
  {P.}~\bibnamefont {Canfield}},\ and\ \bibinfo {author} {\bibfnamefont
  {R.}~\bibnamefont {Prozorov}},\ }\bibfield  {title} {\bibinfo {title}
  {Measuring the lower critical field of superconductors using nitrogen-vacancy
  centers in diamond optical magnetometry},\ }\href
  {https://doi.org/10.1103/PhysRevApplied.11.014035} {\bibfield  {journal}
  {\bibinfo  {journal} {Phys. Rev. Applied}\ }\textbf {\bibinfo {volume}
  {11}},\ \bibinfo {pages} {014035} (\bibinfo {year} {2019})}\BibitemShut
  {NoStop}%
\bibitem [{\citenamefont {Joshi}\ \emph {et~al.}(2020)\citenamefont {Joshi},
  \citenamefont {Nusran}, \citenamefont {Tanatar}, \citenamefont {Cho},
  \citenamefont {Bud'ko}, \citenamefont {Canfield}, \citenamefont {Fernandes},
  \citenamefont {Levchenko},\ and\ \citenamefont {Prozorov}}]{Joshi2020}%
  \BibitemOpen
  \bibfield  {author} {\bibinfo {author} {\bibfnamefont {K.~R.}\ \bibnamefont
  {Joshi}}, \bibinfo {author} {\bibfnamefont {N.~M.}\ \bibnamefont {Nusran}},
  \bibinfo {author} {\bibfnamefont {M.~A.}\ \bibnamefont {Tanatar}}, \bibinfo
  {author} {\bibfnamefont {K.}~\bibnamefont {Cho}}, \bibinfo {author}
  {\bibfnamefont {S.~L.}\ \bibnamefont {Bud'ko}}, \bibinfo {author}
  {\bibfnamefont {P.~C.}\ \bibnamefont {Canfield}}, \bibinfo {author}
  {\bibfnamefont {R.~M.}\ \bibnamefont {Fernandes}}, \bibinfo {author}
  {\bibfnamefont {A.}~\bibnamefont {Levchenko}},\ and\ \bibinfo {author}
  {\bibfnamefont {R.}~\bibnamefont {Prozorov}},\ }\bibfield  {title} {\bibinfo
  {title} {{Quantum phase transition inside the superconducting dome of
  Ba(Fe$_{1-x}$ Co$_ x$)$_2$As$_2$ from diamond-based optical magnetometry}},\
  }\href {https://doi.org/10.1088/1367-2630/ab85a9} {\bibfield  {journal}
  {\bibinfo  {journal} {New J. Phys.}\ }\textbf {\bibinfo {volume} {22}},\
  \bibinfo {pages} {053037} (\bibinfo {year} {2020})}\BibitemShut {NoStop}%
\bibitem [{\citenamefont {Van~Degrift}(1975)}]{VanDegrift1975}%
  \BibitemOpen
  \bibfield  {author} {\bibinfo {author} {\bibfnamefont {C.~T.}\ \bibnamefont
  {Van~Degrift}},\ }\bibfield  {title} {\bibinfo {title} {Tunnel diode
  oscillator for 0.001 ppm measurements at low temperatures},\ }\href
  {https://doi.org/http://dx.doi.org/10.1063/1.1134272} {\bibfield  {journal}
  {\bibinfo  {journal} {Review of Scientific Instruments}\ }\textbf {\bibinfo
  {volume} {46}},\ \bibinfo {pages} {599} (\bibinfo {year} {1975})}\BibitemShut
  {NoStop}%
\bibitem [{\citenamefont {Prozorov}\ \emph {et~al.}(2000)\citenamefont
  {Prozorov}, \citenamefont {Giannetta}, \citenamefont {Carrington},\ and\
  \citenamefont {Araujo-Moreira}}]{Prozorov2000}%
  \BibitemOpen
  \bibfield  {author} {\bibinfo {author} {\bibfnamefont {R.}~\bibnamefont
  {Prozorov}}, \bibinfo {author} {\bibfnamefont {R.~W.}\ \bibnamefont
  {Giannetta}}, \bibinfo {author} {\bibfnamefont {A.}~\bibnamefont
  {Carrington}},\ and\ \bibinfo {author} {\bibfnamefont {F.~M.}\ \bibnamefont
  {Araujo-Moreira}},\ }\bibfield  {title} {\bibinfo {title} {Meissner-london
  state in superconductors of rectangular cross section in a perpendicular
  magnetic field},\ }\href {https://doi.org/10.1103/PhysRevB.62.115} {\bibfield
   {journal} {\bibinfo  {journal} {Phys. Rev. B}\ }\textbf {\bibinfo {volume}
  {62}},\ \bibinfo {pages} {115} (\bibinfo {year} {2000})}\BibitemShut
  {NoStop}%
\bibitem [{\citenamefont {Kim}\ \emph {et~al.}(2018)\citenamefont {Kim},
  \citenamefont {Tanatar},\ and\ \citenamefont {Prozorov}}]{Kim2018}%
  \BibitemOpen
  \bibfield  {author} {\bibinfo {author} {\bibfnamefont {H.}~\bibnamefont
  {Kim}}, \bibinfo {author} {\bibfnamefont {M.~A.}\ \bibnamefont {Tanatar}},\
  and\ \bibinfo {author} {\bibfnamefont {R.}~\bibnamefont {Prozorov}},\
  }\bibfield  {title} {\bibinfo {title} {Tunnel diode resonator for precision
  magnetic susceptibility measurements in a mk temperature range and large dc
  magnetic fields},\ }\href {https://doi.org/10.1063/1.5048008} {\bibfield
  {journal} {\bibinfo  {journal} {Rev. Sci. Instr.}\ }\textbf {\bibinfo
  {volume} {89}},\ \bibinfo {pages} {094704} (\bibinfo {year} {2018})},\
  \Eprint {https://arxiv.org/abs/https://doi.org/10.1063/1.5048008}
  {https://doi.org/10.1063/1.5048008} \BibitemShut {NoStop}%
\bibitem [{\citenamefont {Prozorov}(2021)}]{Prozorov2021}%
  \BibitemOpen
  \bibfield  {author} {\bibinfo {author} {\bibfnamefont {R.}~\bibnamefont
  {Prozorov}},\ }\bibfield  {title} {\bibinfo {title} {{Meissner-London
  susceptibility of superconducting right circular cylinders in an axial
  magnetic field}},\ }\href@noop {} {\bibfield  {journal} {\bibinfo  {journal}
  {Phys. Rev. Appl.}\ }\textbf {\bibinfo {volume} {16}},\ \bibinfo {pages}
  {024014} (\bibinfo {year} {2021})}\BibitemShut {NoStop}%
\bibitem [{\citenamefont {Gaggioli}\ \emph {et~al.}(2022)\citenamefont
  {Gaggioli}, \citenamefont {Blatter},\ and\ \citenamefont
  {Geshkenbein}}]{Gaggioli2022}%
  \BibitemOpen
  \bibfield  {author} {\bibinfo {author} {\bibfnamefont {F.}~\bibnamefont
  {Gaggioli}}, \bibinfo {author} {\bibfnamefont {G.}~\bibnamefont {Blatter}},\
  and\ \bibinfo {author} {\bibfnamefont {V.~B.}\ \bibnamefont {Geshkenbein}},\
  }\bibfield  {title} {\bibinfo {title} {{Creep effects on the Campbell
  response in type-II superconductors}},\ }\href
  {https://doi.org/10.1103/PhysRevResearch.4.013143} {\bibfield  {journal}
  {\bibinfo  {journal} {Phys. Rev. Res.}\ }\textbf {\bibinfo {volume} {4}},\
  \bibinfo {pages} {013143} (\bibinfo {year} {2022})}\BibitemShut {NoStop}%
\bibitem [{\citenamefont {Brandt}(1991)}]{Brandt1991}%
  \BibitemOpen
  \bibfield  {author} {\bibinfo {author} {\bibfnamefont {E.~H.}\ \bibnamefont
  {Brandt}},\ }\bibfield  {title} {\bibinfo {title} {Penetration of magnetic ac
  fields into type-ii superconductors},\ }\href@noop {} {\bibfield  {journal}
  {\bibinfo  {journal} {Phys. Rev. Lett.}\ }\textbf {\bibinfo {volume} {67}},\
  \bibinfo {pages} {2219} (\bibinfo {year} {1991})}\BibitemShut {NoStop}%
\bibitem [{\citenamefont {Prozorov}\ and\ \citenamefont
  {Kogan}(2018)}]{Prozorov2018}%
  \BibitemOpen
  \bibfield  {author} {\bibinfo {author} {\bibfnamefont {R.}~\bibnamefont
  {Prozorov}}\ and\ \bibinfo {author} {\bibfnamefont {V.~G.}\ \bibnamefont
  {Kogan}},\ }\bibfield  {title} {\bibinfo {title} {Effective demagnetizing
  factors of diamagnetic samples of various shapes},\ }\href
  {https://doi.org/10.1103/PhysRevApplied.10.014030} {\bibfield  {journal}
  {\bibinfo  {journal} {Phys. Rev. Applied}\ }\textbf {\bibinfo {volume}
  {10}},\ \bibinfo {pages} {014030} (\bibinfo {year} {2018})}\BibitemShut
  {NoStop}%
\bibitem [{\citenamefont {Tinkham}(2004)}]{Tinkham2004}%
  \BibitemOpen
  \bibfield  {author} {\bibinfo {author} {\bibfnamefont {M.}~\bibnamefont
  {Tinkham}},\ }\href {http://www.worldcat.org/isbn/0486435032} {\emph
  {\bibinfo {title} {Introduction to Superconductivity}}},\ \bibinfo {edition}
  {2nd}\ ed.\ (\bibinfo  {publisher} {Dover Publications},\ \bibinfo {year}
  {2004})\BibitemShut {NoStop}%
\bibitem [{\citenamefont {Prozorov}\ and\ \citenamefont
  {Giannetta}(2006)}]{Prozorov2006}%
  \BibitemOpen
  \bibfield  {author} {\bibinfo {author} {\bibfnamefont {R.}~\bibnamefont
  {Prozorov}}\ and\ \bibinfo {author} {\bibfnamefont {R.~W.}\ \bibnamefont
  {Giannetta}},\ }\bibfield  {title} {\bibinfo {title} {{Magnetic Penetration
  Depth in Unconventional Superconductors}},\ }\href
  {https://doi.org/10.1088/0953-2048/19/8/r01} {\bibfield  {journal} {\bibinfo
  {journal} {Supercond. Sci. Techn.}\ }\textbf {\bibinfo {volume} {19}},\
  \bibinfo {pages} {R41} (\bibinfo {year} {2006})}\BibitemShut {NoStop}%
\bibitem [{\citenamefont {Eilenberger}(1968)}]{Eilenberger1968}%
  \BibitemOpen
  \bibfield  {author} {\bibinfo {author} {\bibfnamefont {G.}~\bibnamefont
  {Eilenberger}},\ }\bibfield  {title} {\bibinfo {title} {{T}ransformation of
  {G}orkov's equation for type {II} superconductors into transport-like
  equations},\ }\href {https://doi.org/10.1007/BF01379803} {\bibfield
  {journal} {\bibinfo  {journal} {Zeitschrift für Physik A Hadrons and
  Nuclei}\ }\textbf {\bibinfo {volume} {214}},\ \bibinfo {pages} {195}
  (\bibinfo {year} {1968})}\BibitemShut {NoStop}%
\bibitem [{\citenamefont {Prozorov}\ and\ \citenamefont
  {Kogan}(2011)}]{Prozorov2011}%
  \BibitemOpen
  \bibfield  {author} {\bibinfo {author} {\bibfnamefont {R.}~\bibnamefont
  {Prozorov}}\ and\ \bibinfo {author} {\bibfnamefont {V.~G.}\ \bibnamefont
  {Kogan}},\ }\bibfield  {title} {\bibinfo {title} {London penetration depth in
  iron-based superconductors},\ }\href@noop {} {\bibfield  {journal} {\bibinfo
  {journal} {Rep. Prog. Phys.}\ }\textbf {\bibinfo {volume} {74}},\ \bibinfo
  {pages} {124505} (\bibinfo {year} {2011})}\BibitemShut {NoStop}%
\bibitem [{\citenamefont {Kogan}(2013)}]{Kogan2013}%
  \BibitemOpen
  \bibfield  {author} {\bibinfo {author} {\bibfnamefont {V.~G.}\ \bibnamefont
  {Kogan}},\ }\bibfield  {title} {\bibinfo {title} {{Homes scaling and BCS}},\
  }\href {https://doi.org/10.1103/PhysRevB.87.220507} {\bibfield  {journal}
  {\bibinfo  {journal} {Phys. Rev. B}\ }\textbf {\bibinfo {volume} {87}},\
  \bibinfo {pages} {220507} (\bibinfo {year} {2013})}\BibitemShut {NoStop}%
\bibitem [{\citenamefont {Prozorov}\ \emph {et~al.}(2007)\citenamefont
  {Prozorov}, \citenamefont {Vannette}, \citenamefont {Samolyuk}, \citenamefont
  {Law}, \citenamefont {Bud'ko},\ and\ \citenamefont
  {Canfield}}]{Prozorov2007}%
  \BibitemOpen
  \bibfield  {author} {\bibinfo {author} {\bibfnamefont {R.}~\bibnamefont
  {Prozorov}}, \bibinfo {author} {\bibfnamefont {M.~D.}\ \bibnamefont
  {Vannette}}, \bibinfo {author} {\bibfnamefont {G.~D.}\ \bibnamefont
  {Samolyuk}}, \bibinfo {author} {\bibfnamefont {S.~A.}\ \bibnamefont {Law}},
  \bibinfo {author} {\bibfnamefont {S.~L.}\ \bibnamefont {Bud'ko}},\ and\
  \bibinfo {author} {\bibfnamefont {P.~C.}\ \bibnamefont {Canfield}},\
  }\bibfield  {title} {\bibinfo {title} {{{C}ontactless measurements of
  {S}hubnikov-de {H}aas oscillations in the magnetically ordered state of
  {C}e{A}g{S}b$_2$ and {S}m{A}g{S}b$_2$~{s}ingle crystals}},\ }\href {<Go to
  ISI>://WOS:000243894600077 http://prb.aps.org/pdf/PRB/v75/i1/e014413}
  {\bibfield  {journal} {\bibinfo  {journal} {Phys. Rev. B}\ }\textbf {\bibinfo
  {volume} {75}},\ \bibinfo {pages} {014413} (\bibinfo {year}
  {2007})}\BibitemShut {NoStop}%
\bibitem [{\citenamefont {Prozorov}\ \emph {et~al.}(2004)\citenamefont
  {Prozorov}, \citenamefont {Lawrie}, \citenamefont {Hetel}, \citenamefont
  {Fournier},\ and\ \citenamefont {Giannetta}}]{Prozorov2004}%
  \BibitemOpen
  \bibfield  {author} {\bibinfo {author} {\bibfnamefont {R.}~\bibnamefont
  {Prozorov}}, \bibinfo {author} {\bibfnamefont {D.~D.}\ \bibnamefont
  {Lawrie}}, \bibinfo {author} {\bibfnamefont {I.}~\bibnamefont {Hetel}},
  \bibinfo {author} {\bibfnamefont {P.}~\bibnamefont {Fournier}},\ and\
  \bibinfo {author} {\bibfnamefont {R.~W.}\ \bibnamefont {Giannetta}},\
  }\bibfield  {title} {\bibinfo {title} {{Field-Dependent Diamagnetic
  Transition in Magnetic Superconductor
  ${\mathrm{S}\mathrm{m}}_{1.85}{\mathrm{C}\mathrm{e}}_{0.15}{\mathrm{C}\mathrm{u}\mathrm{O}}_{4\ensuremath{-}y}$}},\
  }\href {https://doi.org/10.1103/PhysRevLett.93.147001} {\bibfield  {journal}
  {\bibinfo  {journal} {Phys. Rev. Lett.}\ }\textbf {\bibinfo {volume} {93}},\
  \bibinfo {pages} {147001} (\bibinfo {year} {2004})}\BibitemShut {NoStop}%
\bibitem [{\citenamefont {Kim}\ \emph {et~al.}(2013)\citenamefont {Kim},
  \citenamefont {Sung}, \citenamefont {Cho}, \citenamefont {Tanatar},\ and\
  \citenamefont {Prozorov}}]{Kim2013}%
  \BibitemOpen
  \bibfield  {author} {\bibinfo {author} {\bibfnamefont {H.}~\bibnamefont
  {Kim}}, \bibinfo {author} {\bibfnamefont {N.~H.}\ \bibnamefont {Sung}},
  \bibinfo {author} {\bibfnamefont {B.~K.}\ \bibnamefont {Cho}}, \bibinfo
  {author} {\bibfnamefont {M.~A.}\ \bibnamefont {Tanatar}},\ and\ \bibinfo
  {author} {\bibfnamefont {R.}~\bibnamefont {Prozorov}},\ }\bibfield  {title}
  {\bibinfo {title} {{Magnetic penetration depth in single crystals of
  SrPd$_{2}$Ge$_{2}$ superconductor}},\ }\href
  {https://doi.org/10.1103/PhysRevB.87.094515} {\bibfield  {journal} {\bibinfo
  {journal} {Phys. Rev. B}\ }\textbf {\bibinfo {volume} {87}},\ \bibinfo
  {pages} {094515} (\bibinfo {year} {2013})}\BibitemShut {NoStop}%
\bibitem [{\citenamefont {Kim}\ \emph {et~al.}(2021)\citenamefont {Kim},
  \citenamefont {Tanatar}, \citenamefont {Hodovanets}, \citenamefont {Wang},
  \citenamefont {Paglione},\ and\ \citenamefont {Prozorov}}]{hynshoo_2021}%
  \BibitemOpen
  \bibfield  {author} {\bibinfo {author} {\bibfnamefont {H.}~\bibnamefont
  {Kim}}, \bibinfo {author} {\bibfnamefont {M.~A.}\ \bibnamefont {Tanatar}},
  \bibinfo {author} {\bibfnamefont {H.}~\bibnamefont {Hodovanets}}, \bibinfo
  {author} {\bibfnamefont {K.}~\bibnamefont {Wang}}, \bibinfo {author}
  {\bibfnamefont {J.}~\bibnamefont {Paglione}},\ and\ \bibinfo {author}
  {\bibfnamefont {R.}~\bibnamefont {Prozorov}},\ }\bibfield  {title} {\bibinfo
  {title} {{Campbell penetration depth in low carrier density superconductor
  YPtBi}},\ }\href {https://doi.org/10.1103/PhysRevB.104.014510} {\bibfield
  {journal} {\bibinfo  {journal} {Phys. Rev. B}\ }\textbf {\bibinfo {volume}
  {104}},\ \bibinfo {pages} {014510} (\bibinfo {year} {2021})}\BibitemShut
  {NoStop}%
\bibitem [{\citenamefont {Prozorov}\ \emph {et~al.}(2003)\citenamefont
  {Prozorov}, \citenamefont {Giannetta}, \citenamefont {Kameda}, \citenamefont
  {Tamegai}, \citenamefont {Schlueter},\ and\ \citenamefont
  {Fournier}}]{Prozorov2003}%
  \BibitemOpen
  \bibfield  {author} {\bibinfo {author} {\bibfnamefont {R.}~\bibnamefont
  {Prozorov}}, \bibinfo {author} {\bibfnamefont {R.~W.}\ \bibnamefont
  {Giannetta}}, \bibinfo {author} {\bibfnamefont {N.}~\bibnamefont {Kameda}},
  \bibinfo {author} {\bibfnamefont {T.}~\bibnamefont {Tamegai}}, \bibinfo
  {author} {\bibfnamefont {J.~A.}\ \bibnamefont {Schlueter}},\ and\ \bibinfo
  {author} {\bibfnamefont {P.}~\bibnamefont {Fournier}},\ }\bibfield  {title}
  {\bibinfo {title} {Campbell penetration depth of a superconductor in the
  critical state},\ }\href {https://doi.org/10.1103/PhysRevB.67.184501}
  {\bibfield  {journal} {\bibinfo  {journal} {Phys. Rev. B}\ }\textbf {\bibinfo
  {volume} {67}},\ \bibinfo {pages} {184501} (\bibinfo {year}
  {2003})}\BibitemShut {NoStop}%
\end{thebibliography}
%

\end{document}